\documentclass[11pt]{article}
\usepackage[utf8]{inputenc}
\usepackage{amsmath,amsfonts,amssymb}
\usepackage{graphicx}

\usepackage{full page}
\usepackage{authblk}

\newtheorem{definition}{Definition}

\newtheorem{lemma}[definition]{Lemma}
\newtheorem{fact}{Fact}
\newtheorem{theorem}[definition]{Theorem}

\newcommand{\Prob}[1]{\mathbf{P} \left( #1 \right)}

\newcommand{\proof}{\noindent\textit{Proof. }}
\newcommand{\qed}{\hspace{\stretch{1}$\square$}}
\newcommand{\skproof}{\noindent\textit{Sketch of Proof. }}
\newcommand{\ideaproof}{\noindent\textit{Idea of the proof. }}

\newcommand{\edges}{\binom{n}{2}}

\newcommand{\Expe}[1]{\mathbf{E}\left[ #1 \right]}

\newcommand{\sG}{\mathcal{G}}
\newcommand{\sE}{\mathcal{E}}
\newcommand{\sdG}{\mbox{dyn-}\mathcal{G}}

\newcommand{\polylog}{\mbox{\rm{polylog\,}}}

\newcommand{\pup}{p_\uparrow}
\newcommand{\pdw}{p_\downarrow}
\newcommand{\qup}{q_\uparrow}
\newcommand{\qdw}{q_\downarrow}

\newcommand{\orlink}{{\sc link-proc}}

\newcommand{\ErdRen}{Erd\"{o}s-R\'{e}nyi }

\author[1]{Andrea Clementi}
\author[1]{Miriam Di Ianni}
\author[1]{Giorgio Gambosi}
\author[2]{Emanuele   Natale}
\author[3]{Riccardo Silvestri}
 
 \affil[1]{ Universit\`a \emph{Tor Vergata} di Roma, {\tt ``lastname''@mat.uniroma2.it}}
\affil[2]{Master-Degree Student, Universit\`a \emph{Tor Vergata} di Roma,   {\tt emanatale@gmail.com}}
  \affil[3]{ \emph{Sapienza} Universit\`a di Roma,   {\tt silvestri@di.uniroma1.it}  }

\title{\textbf{Distributed  Community Detection in Dynamic Graphs\thanks{
Partially supported by the Italian National Project COFIN-PRIN 2010-11 \emph{ARS TechnoMedia} (Algoritmica per le Reti Sociali Tecno-mediate).} }}

\begin{document}

\maketitle

\begin{abstract}
Inspired by the increasing interest in  self-organizing social opportunistic networks, we investigate the problem
of distributed detection of  unknown communities in dynamic random graphs.  As a formal framework, we consider 
the dynamic version of the well-studied \emph{Planted Bisection Model} $\sdG(n,p,q)$ 
 where the node set $[n]$ of the network  is partitioned into   two unknown
  communities and, at  every  time step, each possible
    edge $(u,v)$ is active    with probability $p$ if both nodes belong to the same community,
  while it is active   with probability $q$ (with $q<<p$) otherwise. We also consider a time-Markovian generalization
   of this model.
  
 We propose   a distributed protocol based 
 on the popular \emph{Label Propagation Algorithm} and prove that, when the ratio $p/q$ is larger than   $n^{b}$ (for an
 arbitrarily small constant $b>0$),
 the protocol finds  the right ``planted'' partition in $O(\log n)$  time even when the snapshots of the dynamic graph are sparse and disconnected (i.e. in 
 the case $p=\Theta(1/n)$).

\end{abstract}


\section{Introduction}

\noindent
Community detection in complex networks has recently attracted wide attention in several research areas such as social networks,
communication networks, biological systems \cite{GN02, EK10}. The general notion of community   refers to the fact that nodes 
tend to form  clusters which are more densely interconnected relatively to the rest of the network. Understanding 
the community structure of a complex network     is a challenging crucial issue in   several applications.
Good surveys  on this topic can be found in \cite{BLMCH06,DDDA05,NG04}.
For instance, in biological networks, it is widely believed that modular structures plays an important role in biological functions \cite{RSMOB02},
while 
in Online Social Networks such as Facebook, community detection is vital  for
the design of related applications, devising business strategies and may even have direct implications
on the design of the network themselves \cite{LN04,FL02}. 
  A    modern 
 application scenario (the one this paper is inspired from) is that of \emph{Opportunistic Networks }
where   recent  studies   show  that \emph{social-aware} protocols provides efficient solutions
for basic communication tasks  \cite{CHCDSG07,SJP08,TL09}. 

The static    \emph{Planted Bisection Model  } \cite{B87,BCLS87,DF89} (or \emph{Stochastic Blockmodel}, as it is known in
  the statistics community \cite{HLL83,SN97})  is a popular framework to formalize the problem of
   detecting  communities in random graphs.
  
  \smallskip
  \noindent
  \textbf{The (Static) Planted Bisection Model: Centralized Algorithms.}
   The (static)  Planted Bisection Model is defined as a static  random graph $\sG(n,p,q)$ (with $p,q \in (0,1)$
  such that $q<<p$) where the node subset
  $[n] = \{ 1, 2, \ldots, n\}$ is partitioned 
  into two      equal-sized \emph{unknown communities} $V_1$ and  $V_2$ and each possible edge $(u,v)$ is included with probability $p$
  if $u$ and $v$ both belong to the same community while it is included with probability $q$ otherwise\footnote{ 
  Observe that when $p=q$ the random graph model is the well-known \ErdRen\ model}. The  goal here is to 
  identify  the unknown partition. 
  
  Dyer and Frieze  \cite{DF89}  
  show that if $p>q$ then the minimum edge-bisection is the one that
  separates the two classes and  then they  derive an      algorithm working in 
   $O(n^3)$ expected time. This bound has been   then improved to $O(n^{2+ \epsilon})$  (for arbitrarily small $\epsilon$)
   by Jerrum and Sorkin \cite{JS98}
  for some range of $p$ and $q$ by using simulated annealing. Further improvements were obtained by Condon and Karp
  \cite{CK01} that show a linear time algorithm for dense graphs and, more recently, by Mossel et Al \cite{MNS12} that provide
    an efficient
  algorithm and some impossibility result for sparse graphs. We emphasize that all the above algorithms are based on 
  centralized,   expensive procedures such as   simulated annealing and spectral-graph computations: all of them 
   require the full knowledge of the graph adjacency matrix and, moreover, they work on  static  graphs only.

 \smallskip
 \noindent
\textbf{Community Detection in Opportunistic Networks.}
 Recent studies  in opportunistic networks   focus on 
   the impact  of   the    \emph{ agent  social behavior }  
       on  some  basic communication tasks such as  routing and broadcasting \cite{CHCDSG07,SJP08,TL09}. 
Recently this issue has been investigated  in an emerging class of opportunistic networks called 
     \emph{Intermittently-Connected
Mobile Networks (ICMNs)}  \cite{WCdeA11}: such networks are   characterized by   wireless  links, representing
 opportunities for exchanging data,
 that sporadically appear  
    among network nodes (usually  mobile radio devices).
    So-called \emph{social-aware} communication protocols rely on the reasonable intuition that, since 
  mobile devices are carried by people who tend to form \emph{communities}, 
  members (i.e. nodes) of the same community 
      are used  to communicate with each other much   more often than nodes  from different communities. 
 Experiments    on  real-data sets     have widely shown that identifying communities
   can strongly help in improving 
  the protocol performances \cite{CHCDSG07,SJP08,TL09}.
  It thus follows that  community detection in   ICMNs is a crucial issue.
  
  As observed above, several  centralized community-detection methods have been proposed in the literature  
   that may result useful for offline data analysis of mobile traces. However, it is a common belief that  next-future technologies will yield a dramatic growth of \emph{self-organizing} ICMNs
  where the network protocols work without relying on any centralized server. In this new communication paradigm, 
  it is required that community detection is performed in a  fully distributed way.
    It turns out that  the above-discussed  centralized algorithms are   not suitable for community detection in  self-organizing dynamic networks
such as ICMNs. 
To this aim, in this paper we consider an algorithmic solution to community detection in ICMNs  that relies on   the epidemic mechanism known
as \emph{Label Propagation Algorithms} \cite{BC09,LHLC09,LM10,RAK07}: this method will be discussed later in the introduction.

 \smallskip
\noindent
 \textbf{The Dynamic Planted Bisection Model.}  
  In order to capture the high dynamicity of ICMNs, we consider the natural dynamic version of the
  $\sG(n,p,q)$ model. A {\em dynamic graph} is a  probabilistic process that describes a graph whose topology changes with time: so it can be represented by 
a sequence 
$\sG = \{ G_t = ([n],E_t) \,:\, t \in \mathbb{N} \}$ of graphs with the same
set $V = [n]$ of nodes, where $G_t$ is the \emph{snapshot} of the dynamic  graph at time step $t$. 

The dynamic version of  the Planted Bisection Model, denoted as $\sdG(n,p,q)$,   consists of a dynamic graph where
     $n$ is the number of nodes  while $p=p(n)$ and
 $q=q(n)$ are the edge-probability functions.  At every time step $t$, each   edge $(u,v)$ is included in $E_t$   with
 probability $p$ if both $u$ and $v$ belong to the same community $V_i$ ($i=1,2$) while it is included with probability $q$ otherwise
 (this model can  also be seen as 
  a non-homogeneous version  of the   
\emph{dynamic Erd\"os--R\'enyi   graph}  model  \cite{AKL08,CHCDSG07}).
  So, the dynamic state (on/off) of an edge over the time  is a random
  variable having Bernoully distribution with parameter $p$ or $q$, respectively.

 This     model   clearly assumes    important simplifications
that may impact   several properties of real  opportunistic networks:
 for instance, we have assumed  
that contacts  between nodes follow Bernoulli  processes, so  the distribution of time between
two contacts of a pair of nodes  follows an  exponential law. Previous experiments
have shown that this assumption holds
only at the timescale of days and weeks \cite{CHCDSG07,KLV07}.
 However, in \cite{CMMD07},  experimental validations  have shown that   some real ICMNs 
 (e.g. those studied in     the Haggle Project \cite{CHCDSG07} and in  the MIT Reality
Mining Project \cite{EP05}) exhibit some  crucial connectivity properties (such as hop diameter) which are
 well-approximated by \emph{sparse} dynamic Erd\"os--R\'enyi   graphs.  
 
  A strong  simplifying assumption in the  dynamic Erd\"os--R\'enyi   graph model is \emph{time independence}:
the graph topology at time $t$ is fully independent from the topology at time $t-1$. 
 \emph{Edge Markovian Evolving Graphs} (in short \emph{edge-MEG}) were first introduced  in \cite{CMMPS08} as  a generalization of 
  the  dynamic Erd\"os--R\'enyi   graph model that captures the strong dependence
between the existence of an edge at a given time step and its
existence at the previous time step.
An edge-MEG is  a  dynamic random graph  $\sG (n,\pup,\pdw,E_0) = \{ G_t = ([n],E_t) \,:\, t \in \mathbb{N} \}$   defined as follows.
 Starting from an
initial random edge set $E_0$, at every time step, every edge changes its
state (existing or not) according to a two-state Markovian process with probabilities $\pup$
and $\pdw$. If an edge exists at time $t$, then, at time
$t+1$, it disappears with probability $\pdw$. If instead the edge does not exist at time $t$, then
it will come up at time $t+1$ with probability $\pup$.
We observe that the setting $\pdw = 1-\pup$ yields a sequence of independent  Erd\"os--R\'enyi random
graphs, i.e., \emph{dynamic Erd\"os--R\'enyi   graphs}, with edge probability $p = \pup$. 
  Edge-MEGs have been   adopted as concrete models for several real    dynamic networks such as 
  faulty networks \cite{CMPS09}, peer-to-peer systems \cite{V11},   mobile ad-hoc networks \cite{V11}, and   vehicular networks
  \cite{MRRS12}.
  Furthermore, Edge-MEGs have  been    considered   by Whitbeck et al \cite{WCdeA11} as a concrete model for analyzing the
  performance of   epidemic routing  on 
  sparse ICMNs and the  obtained theoretical results have been also validated over  real   trace data  
  such as the \emph{Rollernet} traces \cite{TL09}.
 In this paper, we consider the Edge-MEG as a mathematical model for ICMNs.  
   The dynamic Planted Bisection Model can be easily generalized in order to include   edge-MEGs:    here,  
   we have   two   edge-probability parameter pairs $(\pup,\pdw)$ and 
 $(\qup,\qdw)$ between two nodes $u$ and $v$    depending on whether they both belong to the same community or not.
 So, if both $u$ and $v$   belongs to the same community then the edge $(u,v)$ is governed by 
 the 2-state Markov chain with parameters $(\pup,\pdw)$ otherwise the edge is governed by the 2-state Markov chain
 with parameter $(\qup,\qdw)$. We assume that $\qup << \pup$ and, 
 according to the parameter tuning performed  in \cite{WCdeA11}, it turns out that  the best fitting to real scenarios 
 is achieved by setting       $\pdw$, (and $\qdw$) as absolute constants.
 This is mainly due to the fact that, once  a   connection  comes up, its  expected life-time
  does not depend on the size of the network  \cite{WCdeA11}.

    The algorithmic goal in the        $\sdG(n,p,q)$ model  is to design a fully-distributed
    protocol that computes a \emph{good (node) labeling} for the dynamic graph 
   $\sG$: 
  we say that a function  $Z:V\rightarrow \{1,2 \}$ is a \emph{good labeling} for
$\sG$  if $Z$  labels each community with  a different label: 
$\forall i,k\in\{1,2\} \,\forall u\in V_i\,\forall v\in V_k\ : \ Z(u)=Z(v) \ \leftrightarrow \  i=k$.\\
 Nodes  are entities that share 
   a global clock (this is reasonable in opportunistic networks 
    by assuming each node to be  equipped by a GPS) and       know  (a good approximation of)
   the number $n$ of nodes but it is not required they  have distinct IDs. 
   Initially, 
   each node does  not know anything about its own community and it  is not able to distinguish    the community of its  neighbors.
  At every time step, every node can exchange information with      its current  neighbors.

  In \cite{HYCC07},  some     greedy protocols are   tested 
  on specific sets of   real mobility-trace datas. By running such protocols, every node constructs and updates its own community-list
  according to the length and the rate of the contacts observed so far by itself and by the nodes it meets. 
  So, the protocol   exploits the   intuition  that communities are formed by nodes that use
  to meet each other often and for a long time. 
 However, no analytical result is given for  such heuristics that, moreover,   require
nodes   to often    update and transmit  relatively large lists of node-IDs during   all the process:
the resulting overhead  may be  too heavy in several opportunistic networks such as ICMNs.

  \smallskip
  \noindent
 \textbf{Label Propagation Algorithms.}
 A well-studied community-detection strategy is the one  known as \emph{Label Propagation Algorithms (LPA)}  
  \cite{RAK07}. This strategy is based on a simple epidemic mechanism which can be efficiently implemented in 
  a fully-distributed fashion since it  requires easy local computations:  it is thus very suitable
  for opportunistic networks such as ICMNs. In its basic  version,
   some distinct  labels are initially  assigned to a subset of nodes; at every  step, each node updates its label (if any) by choosing the
    label which most of its (current) neighbors have (the \emph{majority} label); if there are multiple majority labels, one label is chosen randomly.
  Clearly, the goal of the protocol is to  converge to a good labeling for $\sdG(n,p,q)$. 
    
   Despite the simplicity of LPA-based protocols, very few analytical results are known on 
   their performance over relevant classes of graphs.  It seems   hard to derive, from   empirical results,
any fundamental conclusions about LPA behavior, even on specific families
of graphs \cite{KPS13}. One reason for this  hardness is
that despite its simplicity, even on simple graphs, LPA can have complex 
behavior, not far from  epidemic processes  such as the  spread of disease in an interacting  population \cite{N02}.  

  Several versions of   LPA-based protocols have been      tested on a wide range of social networks  
  \cite{BC09,CG10,LM10,LHLC09,RAK07}: such works    experimentally 
  show that LPA-based  protocols work quite efficiently and are effective in providing    \emph{almost} good labeling.  
  Based on extensive  simulations, Raghavan et al \cite{RAK07}
  and Leung et al \cite{LHLC09} empirically show that the average convergence
  time of the (synchronous) LPA-based protocols is bounded by some  logarithmic function on $n$.
  Clearly, the goal of the protocol is to  converge to a good labeling for $\sdG(n,p,q)$. 
    Despite the simplicity of LPA-based protocols, very few analytical results are known on 
   their performance over relevant classes of graphs.  As observed in  \cite{KPS13}, it seems   hard to derive, from   empirical results,
any fundamental conclusions about LPA behavior, even on specific families
of graphs. 
Recently,  Cordasco and Gargano \cite{CG12} provided 
  a semi-synchronous version of the LPA-based protocol and  formally  prove 
  that it guarantees finite convergence time  on any static graph.
      In \cite{KPS13}, an LPA-based protocol has been analyzed on the  Planted  Partition Model for highly-dense    topologies.
  In particular,  their analysis considers     the static model $\sG(n,p,q)$ with $p = \Omega(1/n^{1/4-\epsilon})$ and $q = O(p^2)$:  
  observe that in this case
  there are  (w.h.p.)  highly inter-connected communities having constant diameter 
  and a  relatively-small cut among them. In this very restricted case, they show
  the protocol converges in constant expected time and conjectured a logarithmic bound for sparse topologies.

      In general, providing analytical bounds on  the convergence time of LPA-based
  protocols over relevant classes of   networks
   is an important open question that has been proposed in several papers arising from different areas \cite{BC09,CG10,KPS13,
   LHLC09,RAK07}.

 \smallskip
 \noindent
 \textbf{Our Algorithmic Contribution.}
  We provide an  efficient    distributed  LPA-based   protocol    on  the    dynamic Planted Bisection Model
 $\sdG(n,p,q)$    with    arbitrary   $p>0$   and  $q = O(p/ n^b)$ where   $b>0$ is any \emph{ arbitrarily small}   constant.
     Our  protocol yields 
 with high probability\footnote{As usual, we say an event holds with high probability if it holds
 with probability at least  $1-\frac 1{n^{\Theta(1)}}$. } (in short $w.h.p.$) a 
 good labeling  in $O\left(\max\{\log n, \frac{\log n}{pn}\}\right)$ time. The bound is tight for any $p = O(1/n)$
  while it  is
 only a logarithmic factor larger than the optimum for the rest of the parameter range (i.e. for more dense topologies).
 For the first time,  we thus formally prove a  logarithmic bound on the convergence time of an LPA-based protocol
   on a  class of   sparse and disconnected dynamic random graphs (i.e. for $p= \Theta(1/n)$).
 The local labeling  rule adopted by the protocol is  simple and   requires no node IDs: 
  the only exchanged informations are  the  labels.    
 Our protocol can be
  easily adapted in order to construct a good labeling in the presence of a larger 
   number of equal-sized communities (provided that this number is an absolute 
  constant)
  and, more importantly, it also works    for   the Edge-MEG model   
   $\sG (n,\pup,\pdw,\qup,\qdw,E_0)$    in the parameter   range 
  $\qup = O(\pup/n^b)$, where $b$ is \emph{any} positive constant. In the latter model, the completion time is w.h.p. bounded by 
  
  \[  O \left( M \cdot  \max\left\{\log n, \frac{\log n}{\pup n}\right\}   \right)     \]  
  
  \noindent
  where $M$ is a bound on the \emph{mixing time} of the two 2-state Markov chains governing the edges of the dynamic graph.
  It is known that   (see for example \cite{CMMPS08})
  \[ M \ = \  O\left( \max\left\{ \frac1{\pup+\pdw} , \frac1{\qup+\qdw}, \log n \right\} \right) \] 
  Observe that, when  $\pdw$ and $ \qdw$ are some arbitrary positive constants and $\pup = \Omega(1/n)$ (this case includes the
  ``realistic'' range derived in \cite{WCdeA11}), then $M= O(\log n)$ and
  the bound on the completion time becomes $O(\log^2n)$. This bound    is only a  logarithmic factor larger than the optimal
  labeling time  in the case of sparse   topologies, i.e., when $\pup = \Theta(1/n)$.

 We    run our protocol over hundreds  of random instances according to the 
 $\sdG(n,p,q)$ model with $n$ varying from $10^3$ to $10^6$.
 Besides a good validation of our asymptotical analysis, the experiments show further positive features of the  protocol.
 Our protocol is indeed \emph{tolerant}   to   non-homogeneous edge-probability functions.
 In particular, the  protocol almost-always returns  a 
  good labeling in   \emph{Bernoullian} graphs where the edge probability is not uniform, i.e.,
for each  pair $(u,v)$ of nodes in the same community, the parameter  $p_{u,v}$    is  suitably  chosen  in   order to 
         yield        irregular sparse graphs.
    A detailed description of the   experimental results can be found   in the  Appendix (Section \ref{sec:expe}).

 \subsection{A Restricted Setting: Overview}
Let us      consider the  dynamic graph  $\sdG(n,p,q)$ and, for the sake of clarity,
 we first assume the following restrictions hold: the parameter  $p$ is known by every node; there are only 2 communities $V_1$ and $V_2$, each of size $n/2$ ($n$ is an even number);
 the labeling process starts with (exactly) two
 \emph{source nodes},  $s_{1}\in V_{1}$ that is labeled by $z_{1}$ and  
$s_{2}\in V_{2}$ that is labeled by $z_{2}$ with $z_1 \neq z_2$. 
The parameters  $p$ and $q$ belong to the following ranges
\begin{small}
\begin{equation} \label{eq:Cpq}
\frac 1n \ \leqslant \ p \ \leqslant \ \frac{d \log n }{n} \ \mbox{ and }  q = O\left( \frac{p}{n^{b}}\right)
, \mbox{ for some constants  $d>0$ and $b>0$.}
\end{equation}
\end{small}
 Such restricions make the description   easier, thus   allowing us to focus on the main  ideas  of our  protocol and  of   its analysis. Then,
   in Subsection \ref{ssec:general1} and in the Appendix, we will show how to remove the above assumptions  in order to prove
     the general result stated in the introduction.
 
The protocol relies on the    simple and natural properties of  LPA. Starting from two source nodes (one in each community), each one  having
  a different label, the protocol performs
a label spreading by adopting a  simple labeling/broadcasting rule (for instance, 
 every node gets the label it sees most frequently in its neighbors). 
Since links between nodes  of the same community
are much more frequent than the other ones, we  can  argue that the \emph{good-labeling}  will be faster than the \emph{bad-labeling }  
(in each community, the good labeling is the one  from the  source of the community while the bad labeling is the one
coming from the other source).

However, providing a rigorous analysis of the above process requires to cope with some  non-trivial probabilistic issues that 
have not been considered  in  the analysis of information spreading  in   dynamic graphs made in previous papers \cite{BCF09,CMMPS08,CMPS09}.   
Let us consider any local labeling  rule that  depends on the label configuration of the (dynamic) neighborhood of the node only.
 At a given time step, there is  a subset $I_c \subseteq [n]$ of labeled nodes and we need to evaluate
the probabilities $P_g$  ($P_b$) that a non-labeled node gets a  good (bad) label in the next step. After an initial phase,  there is a non-negligible probability that 
some nodes   will get the bad label. Then, such nodes will start  a spreading of  the  bad labeling  
at the same rate of the good one.  Observe also that  good-labeled nodes may (wrongly) change their state
as well, so, differently from a standard single-source broadcast, 
the epidemic process is not monotone with respect to good-labeling.

  It turns out that   the    probabilities $P_g$ and $P_b$  strongly  depend on    the
\emph{label-balance} between the sizes  of the subsets of well-labeled nodes  and of  the badly-labeled ones in the two communities.
Keeping a tight balance  between such values  during all the process is the main technical goal of the protocol.
In  arbitrary label configurations over \emph{sparse} graph snapshots, 
          getting  ``high-probability'' bounds on the
  rate of new (well/badly) labeled nodes is a non-trivial   issue:  indeed,  it is not hard to show
  that, given any two   nodes $v,w \in [n]\setminus I_c$, the events ``$v$ will be (well/badly-)labeled''
and ``$w$ will be (well/badly-)labeled''  are not independent.

\noindent
As we will see, 
 such issues are already present in  the ``restricted'' case   considered in this section.


\noindent
A first important step  of our approach is to describe the combination between the labeling process and
the dynamic graph as a finite-state Markovian process. 
Then, we perform a step-by-step analysis, focusing on the probability that the Markovian Process
visits a sequence of states having ``good-balance''   properties.  

\noindent
Our protocol applies local rules  depending on
the current node's neighborhood and on the current time step only.
The protocol execution over the dynamic graph
can be   represented by the  following \emph{Markovian Process}:  for any time step $t$, we 
 denote as $\left(k_{1}^{(t)},k_{2}^{(t)},h_{1}^{(t)},h_{2}^{(t)}; E_t\right)$ the \emph{state} reached by the  Markovian Process 
where $k_i^{(t)}$ denotes the number of nodes in the $i$-th community labeled by label $z_i$ at time step $t$ and $h_i^{(t)}$ denotes the number of nodes in the $i$-th community labeled by label $z_j$ at time step $t$, for $i,j=1,2$ and $j \neq i$.
In particular, the    Markovian  Process    works
as follows

\[
\ldots \rightarrow \ \left(k_{1}^{(t)},k_{2}^{(t)},h_{1}^{(t)},h_{2}^{(t)}; E_t\right)
\stackrel{ \mbox{ {\tiny $\sdG(n,p,q)$} }}{ \longrightarrow } 
\left(k_{1}^{(t)},k_{2}^{(t)},h_{1}^{(t)},h_{2}^{(t)}; E_{t+1}\right) \stackrel{ \mbox{{\tiny   protocol}}}{ \longrightarrow }   \]
\[ \stackrel{ \mbox{{\tiny   protocol}}}{ \longrightarrow }  
 \left(k_{1}^{(t+1)},k_{2}^{(t+1)},h_{1}^{(t+1)},h_{2}^{(t+1)}; E_{t+1}\right)  \stackrel{ \mbox{ {\tiny $\sdG(n,p,q)$} }}{ \longrightarrow }  \ldots   \]
The main advantage of this   description  is  the following:  observe the process in any fixed state and consider the set of nodes $U$ still having
no label. 
 Then it is not hard to verify that, in the next time step,    the events  $\{$``node $v$ gets a good/bad label'',  $v\in U\}$, are mutually
 independent. This will  allow us to   prove strong-concentration bounds on the     label-balance  discussed    above
  for a sufficiently-long sequence of states visited
 by the Markovian Process, thus getting
  a large fraction of well-labeled nodes in each community within a  short time; this corresponds to a 
  first protocol stage called   \emph{fast spreading} of the good labels.
  
   Unfortunately, this independence property does not hold among labeled nodes of  the same community, let's see why in  the next simple
   scenario. Assume that  the   rule is the   majority one,  consider
    two nodes $u$ and $v$ having the  same label $z$ at time
   $t$,  and assume the event $\sE =$ ``node $u$ will keep     label $z$ at time $t+1$'' holds. Then 
   the    event  ``$(u,v)\in E_t \, |  \, \sE$'' is more likely and, thus, according  to the majority rule,  
   the event ``$v$ gets label $z \,  | \, \sE$'' is more likely as well. This clearly shows a  key-depencence in the label spreading.

 In order to overcome this issue, our protocol   allows every  node to change its  first label-updating  rule   only after a    
   \emph{spreading stage} of suitable length (we will see later this stage  is in fact formed by 3 consecutive   phases): we can thus 
  analyze the spreading of the good labeling   (only) on the current set
 of unlabeled nodes  (where stochastic  independence holds) 
  and  prove that  the process  reaches  a state with a  large number of well-labeled nodes.  After this spreading stage,
 labeled nodes (have to) start to update their labels according to some simple rule that will be discussed later.   We prove that this 
 \emph{saturation phase} 
   has logarithmic  convergence time
    by providing a simple and efficient method to cope with  the above discussed stochastic dependence.

\section{A Restricted Setting: Formal Description} 
The protocol works in $5$ consecutive temporal phases: the  goal of this phase partition is to control the rate of   new labeled nodes  as 
function of the expected values reached by the random variables (r.v.s) $k_{i}^{(t)},h_{i}^{(t)}$ (at the end of each phase).
 Indeed, when such expected values reach some specific thresholds, the protocol and/or its analysis must change accordingly 
 in order to keep the label configuration
well-balanced in the two communities  during all 
the process and to manage the stochastic depencence described above.

\noindent
At any time step $t$, we denote, for each node $v\in V_{i}$, the number of $z_{i}$-labeled neighbors of $v$ as $N_{i}^{v}(t)$, for $i=1,2$.
Given a node $v \in V$, the set of its  neighbors at time $t$ will be denoted as
$\Gamma_t (v)$.
For the sake of brevity, whenever possible we will omit the parameter  $t$ in the above variables and,
 in the proofs, we will only analyze the labeling in  $V_{1}$, the analysis for   $V_{2}$ being the same.

\medskip \noindent {\bf Stage I: Spreading}

\smallskip \noindent \underline{Phase 1:   Source Labeling.}
  The phase runs for  $\tau_{1}=c_1\log n$ time steps, where  $c_1>0$ is an explicit constant that will be fixed later.
In this phase, only the neighbors of the   sources will decide  their label.
The goal is to  reach a state  such that w.h.p. $k_{i} = \Theta(\log n)$ and $h_{i}=0$ ($i=1,2$).
 For any non-source node $v$, the labeling rule is   the following. 
     
\begin{itemize}
\item   Let $i \in \{1,2\}$; $v$  gets label $z_{i}$ if there is a
time step $t \leq\tau_1$  such that $s_i \in \Gamma_t(v)$ and, for $j \neq i$ and for  all $t$ such that $1 \leqslant t \leqslant \tau_1$, it holds that 
 $s_{j} \notin \Gamma_t(v)$; 
 \item  In all other cases, $v$ remains unlabeled. 
\end{itemize}

 \noindent
 In  App. \ref{app:scol}, we will show that, at the end of this phase,
 a node gets the good label with probability $\Theta(p\tau_1)$ and, w.h.p., no node will get the
 bad label. From this fact,  we can prove  the following

\begin{theorem} \label{thm:endtimephase1}   Let $d_1>0$ be any  (sufficiently large) constant. Then, 
a constant $c_1>0$ can be fixed  so that,  at  time step $\tau_{1}=c_1 \log n$
 the Markovian Process  w.h.p. reaches a state such that
\begin{equation} \label{eq:startingCond2}
k_{1}^{(\tau_1)}, k_{2}^{(\tau_1)} \in \left[\frac{d_1}{16}pn\log n,4d_1pn\log n\right] \ \mbox{ and } \ h_{1}^{(\tau_1)},h_{2}^{(\tau_1)}=0  
\end{equation}
\end{theorem}

\smallskip \noindent \underline{Phase 2: Fast Labeling I.}
This phase of the Protocol aims to get an exponential rate of the good-labeling  inside every community    in order to 
reach, in $\tau_2= O(\log n)$ steps,  a state such that the number of well-labeled nodes is bounded by some root of $n$ and
the number of badly-labeled ones is still 0.  
Differently from Phase 1,  unlabeled
nodes can   get a label at every time step according to the following rule:
for $\tau_1 < t \leq \tau_1+\tau_2$, at time step $t$ of Phase 2 every  \emph{unlabeled} node $v$   
\begin{itemize}
\item  gets label $z_{1}$ at time $t+1$ iff ${N_{{1}}^{v}(t) >0}$ and ${N_{{2}}^{v}(t)=0}$, 
\item  gets label $z_{2}$ at time $t+1$ iff ${N_{{2}}^{v}(t)>0}$ and ${N_{{1}}^{v}(t)=0}$, 
\item  remains unlabeled at time $t+1$ otherwise.
 \end{itemize}

\noindent
In the next theorem, we assume that, at time step $\tau_1$ (i.e. at the end of Phase 1),
 the Markovian Process reaches a state satisfying Cond. (\ref{eq:startingCond2}).
 In particular, we   assume  that $k^{\tau_1}_i \geqslant \underline{k}_i^{\tau_1}$, where
 $\underline{k}_i^{\tau_1}  =  \frac{d_1}{16}pn\log n$.
 Thanks to Theorem \ref{thm:endtimephase1}, this event holds w.h.p.
  In what follows,  we will make use of the following function

\begin{small}\begin{equation*}
F(n,k) = 2\max \left\{ \sqrt{\frac{ \log n}{k}}, \frac {\polylog n}{ n^{1-a}} \right\}
\end{equation*} \end{small}

At the end of Phase 2, we can prove the Process w.h.p. satisfies the following properties.

\begin{theorem} \label{thm:endtimephase2} 
For any $\eta>0$,
  constants $a$ and $\phi$ can be
  fixed so that,   at the final step of Phase 2
\begin{small}
\[ \tau_2= \frac{1}{\log\left(1+\left(\frac{np}{2}\right)\right)}\log\left(\frac{n^a}{\phi\log^3 n}\right)	+	{\log^{-1}\left[\left(1+\dfrac{np}{2}\right)\left(1	- F(n,\underline{k}_1^{(\tau_1)})\right)\right]}	\log\left(\frac{\log^3 n}{\underline{k}_1^{(\tau_1)}	}\right)	+	\tau_1, \] 
\end{small}
 it holds w.h.p. that
 
\begin{equation}  \label{eq:phase2thesisk} 
\mbox{for } \  i=1,2, \   n^a \  \leqslant \  k_i^{(\tau_2)} \  \leqslant	\  n^a \log^{\eta} n , \   \mbox{ and } \ 
{h}_i^{(\tau_2)}\ =\ 0 . \    
\end{equation} \end{theorem} 
  \ideaproof (See App. \ref{ph2:proof} for the proof).
For each time step $t$, let $X$ and $Y$ be the number of (new) nodes that get, respectively,
 the good and the bad label   in $V_1$
at time step $t+1$.
 We will  prove the following key-fact:   if 
      $k_i = O(n^a)$ ($i=1,2$) for some   constant $a <1$, then  it holds w.h.p. 
  $X  = [(1\pm o(1)) p n/2] \cdot k_1 $ and $Y =0$ (the same holds for $V_2$). 
  From such  bounds, we can derive the   recursive  equations for $k_i^{(t)}$ yielding
  the bounds stated in the theorem.
  \qed

\smallskip \noindent \underline{Phase 3: Fast Labeling  II.}
In this phase   nodes apply the same  rule  of Phase 2 but  we need to separate  the analysis   from the
previous one since, when the ``well-labeled'' subset gets    size larger than  some  root of $n$, we cannot anymore exploit
the fact that the bad labeling is w.h.p.   not started yet (i.e. $h = 0$). However, we will show that  when the well-labeled sets    get
size $\Theta(n/\polylog n)$, the bad-labeled sets have     still size     bounded by some root of $n$.
 We assume that, at the end of Phase 2, the Markovian Process reaches a state satysfying Cond. (\ref{eq:phase2thesisk})
of  Theorem \ref{thm:endtimephase2}.

 \begin{theorem}\label{thm:phase3}
  For any constant $\eta>0$, constants $a_1<1$ and $\gamma >0$ can be fixed
    so that  at the final time step of Phase 3
\[ \tau_3 =  \frac 1{\log\left(1+\left(\frac{np}2\right)\right)} \log\left( \frac{n^{1-a}} {\gamma \log^{3}{n}}\right)	+	\tau_2 \] 
for $i=1,2$,  it holds w.h.p.  that  
\begin{equation}\label{eq:phase3thesisk} 
\frac{n}{\log^3n} \leqslant k_i^{(\tau_3)} \leqslant \frac{n}{ \log^{3-\eta} n } 
\  \mbox{ and } \ {h}_i^{(\tau_3)}  \leqslant     n^{a_1} 
\end{equation}
\end{theorem}
 \ideaproof (see App. \ref{ph3:proof} for the proof).
Let $X$ and $Y$ be the r.v.s defined in the proof of Theorem \ref{eq:phase2thesisk}. The presence of the bad labeling
changes the bounds we obtain as follows.
At time step $t+1$, as long as $k_i, h_i = O(n/\polylog n)$, we will prove that
$X = [(1\pm o(1))p n/2] \cdot k_1  $ and $Y =  [(1\pm o(1)) (ph_1 + qk_2)] n/2 $.
From the above bounds,  we will determine 
   two time-recursive bounds on the r.v. $k_i^{t}$ and $h_i^{t}$ that hold (w.h.p.) for any $t$ s.t.  $k_i^{t}, h_i^{t} = O(n/\polylog n)$.
 Then, thanks to the hypothesis $q = O(p/n^b)$ and to
  the fact that  the  Markovian Process starts Phase 3 from  a very ``unbalanced'' state ($k_i = \Omega(n^a)$ and $h_i=0$),
   we  apply the recursive bounds and show    that a time step $\tau_3$ exists satysfying
   Eq. \ref{eq:phase3thesisk}. 

  \qed

\noindent
Theorems \ref{thm:endtimephase2} and \ref{thm:phase3}    
  guarantee  a very tight range for the r.v. $k_1$ and $k_2$ at the final step of Phase 2 and 3, respectively.
As we will  see later, this tight balance is crucial for removing the hypothesis on  the existence of the two leaders.

\medskip
\noindent {\bf Stage II: Saturation}
 
\smallskip \noindent \underline{Phase 4: Controlled Saturation.}
At the end of Phase 3, the Markovian  Process w.h.p.  reaches a state that satisfies the properties stated in Theorem
\ref{thm:phase3}. The goal of     Phase 4  is to obtain  a (large) constant fraction $\alpha$ (say, $\alpha = 3/4$) of   the
nodes of each community that  get the  good label and, at the same time, to ensure that the  number of bad-labeled nodes 
is  still bounded by some root of $n$.   We cannot guarantee this goal by applying the same labeling rule
of the previous phase: the number of bad-labeled nodes would increase too fast.
 The protocol thus performs    a much ``weaker'' labeling rule
 that is enough for the good labeling while keeping the  final 
 number of    bad-labeled nodes bounded by some root of $n$.

\noindent
The fourth phase consists of three consecutive identical time-windows
during which \emph{every} (labeled or not) node $v \in V$ applies the following simple rule:

\smallskip
\noindent
\emph{Time Window of Phase 4.}\\
For any $t \in [1,T_4 = c_4 \log n]$, $v$ looks at the labels of its neighbors at time $t$ and:
\begin{itemize}
\item If $v$ sees only one label (say, $z$)  for \emph{all} the window time steps, then $v$ gets label $z$;
\item In all the other cases (either $v$ sees more   labels or $v$ does not see any label), $v$
 either keeps its label (if any) or it remains unlabeled.
\end{itemize}
\smallskip

\begin{figure}[!ht]
\centering
\includegraphics[scale=1]{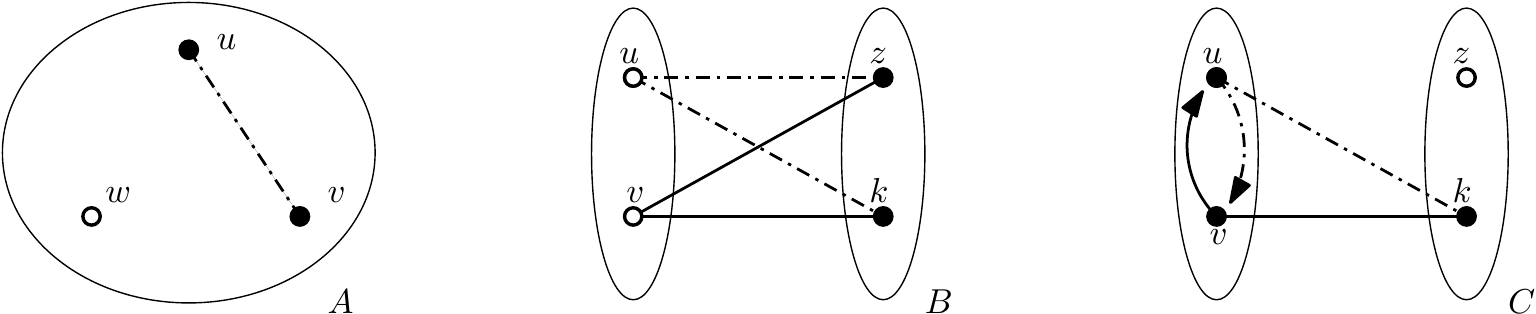}

\caption{ {\small Nodes in  the same community are circled, and they are labeled either  with white  or with black. In graph  $A$ we see that the
 event ``$u$ gets color black'' implies the existence of   edge $(u,v)$, then  $\Prob{\mbox{$u$ gets color black} \,\middle|\, \mbox{$v$ gets color black}} \neq \Prob{\mbox{$u$ gets color black}}$. In  graph $B$ we clearly see that, since $v$ and $u$ does not share any edge toward the other community, these edges do not yield   stochastic dependence. In graph $C$,  we see that if the edges inside each community are directed,
 then the presence  of   bold-drawn edges do not affect the presence of   dashed-drawn edges.}}

\label{fig:stochastic_dep}
\end{figure}

\noindent
\emph{Remark.} Observe that, departing from the previous phases, we now need to analyze the label-spreading of the  above  rule    over nodes
having   previously-assigned   labels. 
This rises   the following   stochastic dependence.
The    analysis  of the previous phases relies on  the independence of the random variables (r.v.s)  that correspond to the events
 ``$u$ gets label $z_i$'' for every label $i$ and every  $u$ in a fixed community $\mathcal{V}$: 
let's enumerate  such  r.v.s   as $\{ X_u \, | \, u \in \mathcal V \}$. Given a node $u$ and a set of nodes $S$,   $E(u,S)$ denotes the set of edges from $u$ to any node in $S$. The  r.v.   $X_u$  depends on the edges incident to $u$; so,  for any  pair $u,v$ 
we can write $X_u = X_u((u,v),E(u,\mathcal{V}\setminus\{v\}), E(u,V\setminus \mathcal{V}))$ 
and $X_v= X_v((v,u),E(v,\mathcal{V}\setminus\{u\}), E(v,V\setminus \mathcal{V}))$. 
 Since in our undirected-graph model  $(u,v)$ equals  $(v,u)$  then  $X_u$ and $X_v$ share the  argument $(u,v)$:
 this clearly yield    stochastic dependence between them (see Fig. \ref{fig:stochastic_dep}). However, if  the graph of $\mathcal{V}$ is  made  directed,
  they become functions of disjoint sets of edges, therefore   $X_1,...,X_{|\mathcal{V}|}$ become  mutually  independent. 
 In order to make our graph directed,   the nodes run a   simple procedure  \orlink\  at the very beginning  of every step.
 This procedure   
  simulates a virtual  $\sdG(n,p,q)$ where   the edges inside each  community $\mathcal{V}$ are  generated according to 
  a directed $G_{n/2, \tilde{p}}$ model, 
  where $\tilde{p}=1-\sqrt{1-p}$. Moreover, the procedure  makes  the resulting probability of the edges between communities   still  bounded by  $O(q)$:  it thus    preserves the polynomial gap between $p$ and  $q$. 
The proofs of these facts are given in   App. \ref{app:stocdep}.

 \smallskip
 \begin{small}
 \noindent
 \textbf{Procedure \orlink:  } 
\begin{enumerate}
\item Each node $u$, for each neighbor $v$ generates a pair $(M^{u}(v), C^{u}(v))$ where $M^{u}(v)$ is a  
 an integer randomly sampled from $[n^3]$, and $C^{u}(v)$ is $1$ or $-1$, each
  with probability $\frac{\sqrt{1-p}-1+p}{p}$ and $0$ with probability $\frac{(1-\sqrt{1-p})^2}{p}$. 
\item $u$ sends this pair to $v$ (so it  receives from $v$  the  pair $(M^{v}(u), C^{v}(u))$).
\item If $M^{u}(v)>M^{v}(u)$ then $u$ defines $D^{u}(v)=C^{u}(v)$, otherwise if $M^{u}(v)<M^{v}(u)$  then  $D^{u}(v)=-C^{v}(u)$. 
\item Finally, $v$ is a (directed) neighbor of $u$   iff $D^{u}(v)\not =-1$.
\end{enumerate}

\end{small}

 \noindent Observe  that we can  neglect the event $M^{u}(v)=M^{v}(u)$ since its  probability  
  is $\frac 1{n^3}$:  if this  happens we   assume that both nodes virtually remove each other from their own neighborhood.
In the sequel, we implicitly assume  that   nodes apply  Procedure \orlink\  and 
the    Protocol-Window of Phase 4  is repeated 3 times for a specific setting of the constant $c_4$ that will be determined in
(the proof of)  Theorem \ref{thm:kh4}.
 Thanks to      Theorem \ref{thm:phase3},  we can assume that 
  the Markovian Process  w.h.p. terminates Phase 3   reaching   a state
 that satisfies Eq. \ref{eq:phase3thesisk}.
    The  proof of the next theorem is given 
  in App. \ref{ph4:proof}.
 
 \begin{theorem} \label{thm:kh4}
Let $\alpha$ be any constant such that $0 < \alpha < 1$. 
Then,
 constants $c_4$ and $a_1<1$  can be fixed so that,  at time step $\tau_4 = \tau_3 + 3 T_4$, the Markovian Process w.h.p reaches  
a state such that, for $i=1,2$,
 
  \begin{equation}\label{eq::f4final}
   k_i^{\tau_4} \ \geqslant \ \alpha n  \  ,  \mbox{ and }
     h_i^{\tau_4} \ \leqslant \  n^{a_1} \ \polylog   n .  
     \end{equation}
      \end{theorem}

\smallskip \noindent \underline{\bf Phase 5: Majority Rule.}
Theorem \ref{thm:kh4} states  that, at the end of Phase 4, the Markovian  Process w.h.p. reaches a state  where 
 a (large) constant fraction of the nodes (say, $3/4$) in both communities  is well-labeled while only 
 $O( n^{a_1} \polylog n)$
nodes are bad-labeled. 
We now  show that a further final phase, where  nodes apply a simple majority rule, yields  the good labeling, w.h.p..
Remind that every node also applies Procedure \orlink\ shown in the previous phase.
\emph{Every} node $v \in V$ applies the following labeling rule:

\begin{itemize}
\item 
For every  $t\in [1,T_5 = c_5 \log n]$, \emph{every} node $v$ observes the labels of its neighbors at time $t$ and,
for every label  $z_i$ ($i=1,2$), $v$ computes the number $f^t_i$ of its neighbors labeled with $z_i$.

\item Then, node $v$ gets label $z_1$ if     $ \sum_{t \in [1,\ldots, \tau_5]}   f^t_1  >   \sum_{ t \in [1,\ldots, \tau_5]} f^t_2$,    
otherwise $v$ gets label $z_2$ (break ties arbitrarily).  
\end{itemize}

\noindent
Let us assume the Markovian Process starts Phase 5  from 
 a state satisfying Eq. \ref{eq::f4final} (say with constant $\alpha = 3/4$).
 The proof of the next theorem is given in App. \ref{ph5:proof}.
 
\begin{theorem} \label{thm:F5all}
 A constant $c_5>0$ can be fixed so    that,
  at time  $\tau_5 = \tau_4 + c_5 \log n$, every  node of each community is well-labeled, w.h.p.
\end{theorem}

\medskip \noindent {\bf Overall Completion Time of the Protocol and its Optimality}

\noindent
When $p$ and $q$ satisfy Cond. (\ref{eq:Cpq}), we have shown 
that every phase   has   length $O(\log n)$:  the Protocol has thus  an overall     completion time $O(\log n)$.  
In Appendix \ref{sec::generalI}, we will show that for $p=o(1/n)$ the length of each phase must be stretched to 
   $\Theta\left(\frac{\log n}{pn}\right)$.
 It is easy to verify that,  if   $p = O(1/n)$, starting from the  initial random snapshot,   there is 
 a non-negligible probability that some  node will be isolated  for  $\tau(n)$ time steps where $\tau(n) $ is any increasing function
 such that $\tau = o\left( \frac{\log n}{pn}\right)$: this implies that, in the above range, our protocol has optimal
 completion time.
 
\section{The General Setting} \label{ssec:general1}

\noindent
 \textbf{Removing the Presence of the  Two Source Nodes.} 
So far we have assumed that, in the initial state of the labeling process, there are exactly two source
nodes,   one in each community, which are labeled with different labels.
This assumption can be removed by introducing a preliminary phase in which a randomized source election is performed and 
by some further changes that are    described below.

\noindent
 In the first step, every node, by an independent random choice,    becomes a \emph{source} with probability 
$\frac{d  \log n}{n}$ for a suitable constant $d>0$.
This clearly guarantees that, in every community, there are w.h.p.  $\Theta(\log n)$ sources.
Then, every source $s_i$ randomly chooses   a label $z_i \in [n^2]$.  This implies that the
minimal label $z_1$ in the first community and the minimal label $z_2$ in the second community are different w.h.p..
Let $a$ and $b$ be the number of sources chosen in $V_1$ and $V_2$, respectively, and define $\ell = a + b$.
We summarize the above arguments in the following 
 
 \begin{fact} \label{fc:n.sources}
 Two positive constants $\eta_1 < \eta_2$ exist such that at the end of the first step w.h.p. it holds  that
 $\eta_1 \log n  \leqslant a,b  \leqslant  \eta_2 \log n \ \mbox{ and } z_1 \neq z_2$.
 \end{fact}
 
 \noindent
 The generic state of the modified Markovian Process is represented by the following set of r.v.s:
 
 \[    (\ell_1,\ell_2; k^1_1,\ldots,k^1_{a},h^1_1, \ldots , h^1_b, k^2_1,  \ldots, k^2_b, h^2_1, \ldots, h^2_a) \]
 where $k^i_j$ equals the number of nodes in $V_i$ labeled by the same (\emph{good}) label as the $j$th source of $V_i$ while $h^i_j$ equals the number
 of nodes in $V_i$ labeled by the same (\emph{bad}) label as the $j$th source of $V_r$ with $r \neq i$. 
At every time step $t$, for any $ v\in [n]$ we define    the r.v. $N_{j}^{v}(t)$  
as the the number of  $v$-neighbors  labeled with label $z_j$ at time $t$.

\noindent
The first three phases of the Protocol are identical to the 2-source case since  
the impact of the presence of an $O(\log n)$ labels in each of the two communities remains  negligible
till the overall number of labeled nodes in each community  is $O(n/ \log^4n)$.
By applying the same analysis of the 2-source case, at the end of Phase 3, we can thus show 
that  the Markovian Process w.h.p. reaches a state having similar properties to those stated in
Theorem \ref{thm:phase3}.
We remind that  
$p$ and $q$ belong to the   ranges in Cond. (\ref{eq:Cpq}).

\begin{theorem} \label{thm:F3multi}
  We can choose a suitable $\tau_3 = \tau_2 + (c_3+o(1)) \log n$ so that, at the end of Phase 3,
 the Markovian Process w.h.p.  reaches a state  in which  for $\ell=1,2$ it holds  
\begin{align*}
\forall j \in [a] \ \ \frac{n}{\log^4n} \ \leqslant \  k^{\ell}_j \ \leqslant \  \frac n{\log^{4-\eta}n} \ ;  \ 
\forall i \in[b]\ \ h^{\ell}_{i}\  = \ \sqrt n \ \polylog n
\end{align*}

\noindent
where $\eta$ is a  constant that can be made arbitrarily small.
\end{theorem}

\noindent
We need to stop at a ``saturation size''
$O(n/\log^{4-\eta}n)$ for \emph{every} good label, since we want     to guarantee (w.h.p.) that 
  the minimal label   infects at least $n/\polylog n$ nodes.
Then, as in the 2-source case, the protocol starts a controlled saturation phase (i.e. Phase 4) 
that consists of (at most) 4 consecutive time-windows in which every node applies the same following 
\emph{minimal-label} rule:

\begin{verse}
For $t=1$ to $T_4 = c_4 \log n$ time steps, $v$ observes the labels of its neighbors and gets
the \emph{minimal} label $\hat z$ among all the observed labels.
\end{verse}

\smallskip
\noindent
Thanks to the above rule, the size of the nodes labeled by the minimal good-label increases by a logarithmic
factor at the end of each of the four windows. This fact can be proved by using  the same arguments of the proof
of Theorem \ref{thm:kh4}. 

\noindent
It thus follows that,  at the end of Phase 4, the number of nodes labeled with  the good minimal
label  is at least a constant (say $3/4$) fraction of all the nodes of the community.
Then, as  in the 2-source case, every node can apply the majority rule in order
to get the right label w.h.p.

     \smallskip
  \noindent
  \textbf{The  Case   \emph{$p$-unknown}.} 
  Our protocol relies on the fact that nodes know the parameter $p=\frac dn$: the length of the protocol's phases
  are functions of $p$.  So an interesting issue is to consider  the scenario
  where  nodes do not know  the parameter $p$ (i.e. the expected degree).
 Thanks to edge independence, the dynamic random-graph process can be seen by every node as an independent
 sequence of random samples. Indeed, at every time step $t$, every node can store   the number $|N^v(t)| $ of its neighbors
 and it knows that this number has been selected by $n-1$ independent experiments
 according to the same  Bernoulli distribution with success probability  $p = \frac dn$. The goal is thus to use such samples
 in order to get  a good
 approximation of $p$. 
    If $p \geqslant \frac 1n$, by using a standard statistical  argument, every node w.h.p.  will get 
    the value of $p$ up to some negligible factor in $O(\log n)$ time. Let's see this task  more formally.
    
    \noindent
     For $c\log n$ time steps (where $c$ is a constant that will be fixed later),
      every node    stores the values  $|N^v(1)|,|N^v(2)|,...,|N^v(c\log n)|$; then it computes
 $S=|N^v(1)|+  \ldots +|N^v(c\log n)|$. 
Since $S$ is the sum of $c\log n \cdot (n-1)$ Bernoulli r.v.s of parameter $\frac{d}{n}$, 
we get  a binomial distribution with mean 
$\Expe{S}=d c\log n  \left(1-\frac{1}{n}\right)$ 
Then, every    node uses the estimator $D(S)=\frac{S}{c \log n  \ \left(1-\frac{1}{n}\right)}$ to guess $d$. 
We can use the Chernoff bound in order to determine  a confidence interval for $D(S)$, as follows 
\begin{small}
\begin{align*}
\Prob{d \notin  \ [D(S)-\delta,D(S)+\delta]}
 =  \Prob{S  \notin [\Expe{S}-\delta c\log n \  \frac{n-1}{n}  ,  \Expe{S}+\delta c \log n \  \frac{n-1}{n} }  = \\
  \Prob{S<\Expe{S}\left(1-\frac{\delta}{d}\right)}+\Prob{S>\Expe{S}\left(1+\frac{\delta}{d}\right)} 
  <  e^{-\frac{\delta^2}{2d^2}\Expe{S}}  +  e^{-\frac{\delta^2}{3d^2}\Expe{S}}    <  4 \left(\frac{1}{n^c}\right)^{\frac{\delta^2}{3d}  }
\end{align*}
\end{small}

\noindent
It thus follows that, for any $d \geqslant 1$, we can choose  $\delta =   \sqrt d$ and $c$ sufficiently large in order to 
get a good confidence interval for all nodes of the network.
This obtained approximation   suffices  to perform an    analysis of the protocol which is equivalent
 to that of the 
case  \emph{$p$-known}.

 \smallskip
  \noindent
  \textbf{More Communities.}  The  presence of a constant number $r = \Theta(1)$ of unknown equally-sized 
   communities can be 
  managed  with  a similar method  to that described above for removing the presence of leaders.
  Indeed,
 the major issue to cope with is the presence of a constant number of different label spreadings in each community and
  the protocol must select the right one in every community. However,  if $r$ is a constant
   and  the number of nodes in each community is some constant fraction of $n$, then 
   the impact of the presence of $O(\log n)$ labels in each of the  $r$ communities remains  negligible
till the overall number of labeled nodes in each community  is $O(n/ \log^4n)$. As in the previous paragraph,
by first applying  the minimal-label rule and then the majority one, the modified protocol returns a good-labeling w.h.p.

Due to lack of space, the protocol analysis in the Edge-MEG model is given in Appendix.
 
\section{Conclusions}
This paper introduces a   framework that allows an analytical  study of 
     the distributed community-detection problem in dynamic graphs. Then, it shows an  efficient algorithmic solution
      in two classes of such graphs that model some features of opportunistic networks such as ICMNs.
     We believe that the problem deserves to be    studied 
in other classes of dynamic graphs that may capture further relevant features of
social opportunistic networks such as geometric constraints.

\smallskip
\noindent
\textbf{Acknowledgements.} We thank Stefano Leucci for its   help in  getting  
 an  
efficient protocol simulation over large random graphs.

 \newpage
\appendix

\newpage
\appendix

\section{More General Settings (Part II)} \label{sec::generalI}
 In this section, we show  the further  relevant generalizations  
that can be efficiently solved by   simple adaptations of our protocol and/or its analysis.

     \smallskip
\noindent
\textbf{Edge Markovian Evolving Graphs.}
 Let us consider an Edge-MEG     $\sdG (n,\pup,\pdw,\qup,\qdw,E_0)$    defined in the introduction
 and assume that $\qup \leqslant \pup/n$. 
  If $0 <  \pup,\pdw,\qup,\qdw < 1$, it is easy to see \cite{CMPS09} that the (unique) stationary distribution of the two corresponding 2-state
    edge-Markov chains (inside and outside the communities, respectively)   are  
\begin{equation*}
\mathbf{\pi^{in}} \ = \  \left( \frac{\pdw}{\pup+\pdw}, \frac{\pup}{\pup+\pdw} \right) \ \mbox{ and } \
 \mathbf{\pi^{out}} \ = \  \left( \frac{\qdw}{\qup+\qdw}, \frac{\qup}{\qup+\qdw} \right) 
\end{equation*}
It thus follows that the dynamic graph, starting from any $E_0$, converges to the  (2-communities)
Erd\"os-R\'enyi random graph  with edge-probability functions \\
\begin{equation*} 
\tilde{p} \ =  \  \frac{\pup}{\pup +\pdw} \ \mbox{ (inside communities) } \ \mbox{ and } \tilde{q} \ =  \  
\frac{\qup}{\qup +\qdw}  \ \mbox{ (outside communities)} \end{equation*} \\
The mixing time $M^{in}$ and $M^{out}$  of the two edge Markov chain are  bounded  by \cite{CMPS09}
 $ M^{in} \ = \ O\left(\frac 1 {\pup + \pdw}\right)$  ,       $M^{out} \ = \ O\left(\frac 1 {\qup + \qdw}\right)$.
Let us  observe that there is a \emph{Markovian dependence} between graphs of consecutive time steps. 
If we observe any  event at time $t$ related to $E_t$ (such as the number of well-labeled nodes) then
  $E_{t+1}$ is not anymore random with the stationary distribution.
  
  \noindent
   It thus follows that we 
  need to change the way the protocol works over  the dynamic random graph.
  Let $M = \max\{M^{in},M^{out}, \log n   \}$;  then by definition of mixing time,   
  starting from \emph{any} edge subset $E_t$ at time $t$,  at   time $t+\Delta$ with some  $\Delta = \Theta(M)$,  if
    $u,v \in V_1$ or $u,v \in V_2$ then   edge $(u,v)$   exists
  with probability $\tilde p   \pm \frac 1{n^2}$, otherwise it exists  with probability $\tilde q   \pm \frac 1{n^2}$.
  In other words, whathever the state of the labeling process is at time $t$, after a time window  proportional to the mixing time,
  the dynamic graph is random with a distribution which is very close to the stationary one.
  We can thus modify our protocol for the \emph{dynamic Erd\"os--R\'enyi   graph} model      $\sdG(n,p,q)$ in order to ``wait for   mixing''.
  Between any two consecutive  steps of the original protocol  there is a \emph{quiescent} time-window of length $\Theta(M)$ where
  every  node simply  does nothing.  Then, the analysis of the protocol over $\sdG (n,\pup,\pdw,\qup,\qdw,E_0)$ is similar to 
   that  in Section \ref{sec::2-source} working for the \emph{dynamic Erd\"os--R\'enyi   graph}        $\sdG(n,\tilde p,\tilde q)$.
   We can thus state that, under the condition  $\qup \leqslant O(\pup/n^b)$ for some constant $b>0$,
    this version of our protocol w.h.p. performs 
   a  good-labeling in time
   $O\left(M \cdot \max\left\{\log n ,\frac{ \log n}{pn} \right\} \right)$.    
We finally observe that, for the ``realistic'' case $\pdw,\qdw = \Theta(1)$ (see the discussion
in the Introduction), the mixing-time bound $M$ turns out to be
$O(\log n)$: we thus get only a logarithmic slowdown-factor w.r.t. the good-labeling   in 
the  \emph{dynamic Erd\"os--R\'enyi   graph}        $\sdG(n,\tilde p,\tilde q)$.

\medskip
  \noindent
  \textbf{Sparse Graphs.} 
When  $p=o\left(\frac{1}{n}\right)$ and $\frac{q}{p}=O\left(\frac{1}{n^b}\right)$ (for some
constant $b>0$), the snapshots of the dynamic graph are very sparse.
So, every node must wait   at least  $\Theta\left(\frac 1{pn}\right)$ time step (in average) in order to meet some other node.
This implies that the labeling protocol will be slower. We can reduce this case to the case $p = 1/n$ by considering 
the \emph{time-union} random graph obtained from    $\sdG(n,p,q)$ according to the following 
  
\begin{definition}
  Let $\Delta$ be any positive integer and  consider any sequence of graphs  $G(V,E_1), \ldots, G(V,E_{\Delta})$.
  Then,  
  we  define  the $\Delta$-OR-graph
\begin{equation*}
G_{\vee}^{\Delta} = \left(V,E^{\Delta} \right) \ \text{ where } \ E^{\Delta} \ = \ \{e \in \edges \  | \ 
\exists t^{\star}\in (1, \Delta] \  : \  e \in E_{ t^{\star}}\}
\end{equation*}  
\end{definition}

\noindent
It is easy to prove the following 

\begin{lemma}\label{lem:orgraph}
Let $p < \frac 1n$, then the $\frac{1}{pn}$-OR-graph of any finite sequence of   graphs selected according  to  the  $\sdG(n, p, q)$ model
is a $\sdG(n,\tilde p,\tilde q)$ with $\tilde p=\Theta\left(\frac{1}{n}\right)$ and $\tilde q=O\left( \frac{p}{n^b} \right)$.
\end{lemma}
 
  \noindent
The modified protocol just works as it would work over 
$\sdG(n,\tilde p,\tilde q)$ with $\tilde p= \Theta\left(\frac{1}{n} \right)$ and $\tilde q=O \left( \frac{p}{n^b} \right)$:
in every phase, every node applies the phase's labeling rule (only)   every $\Delta = \frac 1{pn}$ time steps on the $\frac{1}{pn}$-OR-graph.
  The modified  protocol thus  requires  $\Theta\left(\Delta \log n\right)=\Theta\left(\frac{\log n}{pn}\right)$ time.

   \medskip
  \noindent
  \textbf{Dense Graphs.} 
  When $p$ becomes larger than $\log n/n$ and $q=\left( \frac{p}{n^b} \right)$,
      the labeling problem becomes an easier task since
  standard probability arguments easily  show that 
  the (good) labeling process is faster and the related r.v.s (i.e. number of new labeled nodes 
  at every time step) have much smaller variance.
  This implies that the protocol can be   simplified: for instance, the source-labeling phase (i.e.
  Phase 1) can be 
  skipped while the length of other phases can be reduced significantly as a function of $p$.
  However, we again emphasize that \emph{dense} dynamic random random graphs are not a good model
  for the  scenario we are inspired from:  ICMNs are opportunistis networks having sparse and disconnected topology.
  
  
  \section{Experimental Results} \label{sec:expe}
  We run our protocol over sequences of independent random graphs according to the $\sdG(n,p,q)$ model.
  The protocol has been suitably simplified and tuned in order to optimize the real performance.
  In particular, the implemented   procotol consists of 5 Phases: Phase 1 (Source-Coloring), Phase 2-3 (Fast-Coloring I-II), 
  Phase 4 (Min-Coloring), and   Phase  5 (Majority-Rule). 
  The rules of each phase is the same of the corresponding phase
  analyzed in Section \ref{sec::2-source}. Moreover, the length of every phase is fixed to  $c \log n$. 
  As shown in the next tables,  parameter $c$ is always very small and it depends on  the  parameter $q$.
  The parameter $c$ has been heuristically chosen as the minimal one yielding  the good labeling in more than
  $98 \%$ of the 
  trials. 
  We consider instances of   increasing size $n$ and for each size, we tested 100 random graphs.
  In the first experiment class (see Table 1), we consider homogeneous sparse graphs with the following setting:
  $p= \frac 5n$ and   3  values of $q$ ranging from $1/n^2$ to $1/n^{3/2}$.
   
\begin{table} 
\centering
\caption{Tab. 1. Experimental results for  the homogeneous  case. For every value of $n$, the rows indicates the percentage  of
 good-labeling   for three choices of $q$ and   the ``minimal'' setting for  $c$ (the total number of Protocol' steps is 
 inside   brackets).}  
\begin{tabular}{l|c|c|c|}
n	& $q=n^{-\frac 32},c= 0.9$ 	& $q=n^{-\frac 53},c= 0.6$ 	& $q=n^{-2},c= 0.5$ \\
\hline
20000	&  99 (66) 	& 100 (46) 	& 100 (36) 	\\
40000	&  99 (71) 	& 100 (46) 	& 100 (41) 	\\
80000	& 100 (76) 	& 100 (51) 	& 100 (41) 	\\
160000	& 100 (81) 	& 100 (51) 	& 100 (46) 	\\
320000	& 100 (86) 	& 100 (56) 	&  99 (46) 	\\
640000	& 100 (91) 	& 100 (61) 	& 100 (51) 	\\
1280000	& 100 (91) 	& 100 (61) 	& 100 (51) 	\\
2560000	& 100 (96) 	& 100 (66)	& 100 (56) 	
\end{tabular}
\end{table}
  
  The second class of experiments concerns non-homogeneous random graphs.
  For each pair of nodes $e=(u,v)$ in the same community, the probability $p_e$ is randomly 
  fixed  in a range $[d_1/n, d_2/n]$ before starting the graph-sequence generation. Then, 
 a every  time step  $t \geqslant 0$,   the    graph-snapshot $G(V,E_t)$  is 
  generated by selecting every edge  $e=(u,v)$ according to its birth-probability $p_e$ (the edges 
  between the two communities are generated with parameter $q$).  In Table 2, 
   the experimental results are shown for the case $d_1=1$ and $d_2=9$ in order to generate
   sparse topologies inside the communities, while in Table 3, the   results concern the  
   more dense case where $d_1 = 0$ and $d_2 = \log n$.
   The protocol's implementation is the same of   the homogeneous case above.


\begin{table}
\centering
\caption{Tab. 2.  Experimental results for the non-homogeneous sparse case with $d_1 =1$ and $d_2= 9$.}
\begin{tabular}{l|c|c|c|}
$n$ 	&  $q=1/n^{3/2},c=1$ 	 & $q=1/n^{5/3},c= 0.4$ 	 & $q=1/n^2,c= 0.4$   \\
\hline
20000	&	 100  (46)  &  	 100 (46) 	 & 100 (36) \\ 
40000	&	 98 (71) 	&  99 (46) &  	 100 (41) \\ 
80000	&	 100 (76) & 	 100 (51) &  	 100 (41)  \\
160000	&	 100 (81)  &	 100 (51) 	&  100 (46)  \\
320000	&	 100 (86) &  	 100 (56) 	&  100 (46) \\
640000	&	 100 (91) 	&  100 (61) & 	 100 (51)  \\
1280000	&	 100 (91) 	&  100 (61) &  	 100 (51) 
\end{tabular}
\end{table}

\begin{table}
\centering
\caption{Tab. 3. Experimental results for the non-homogeneous case with $d_1=0$ and $d_2 = \log n$.}
\begin{tabular}{l|c|c|c|}
$n$	& $q=n^{-\frac 32},c=1$	& $q=n^{-\frac 53},c= 0.4$ 	& $q=n^{-2},c= 0.4$ 	\\
\hline
20000	&  99 (76) 	& 100 (31) 	& 100 (31) 	\\
40000	&  99 (81) 	& 100 (31) 	& 100 (31) 	\\
80000	&  98 (86) 	& 100 (31) 	& 100 (31) 	\\
160000	& 100 (91) 	& 100 (36) 	& 100 (36) 	\\
320000	& 100 (96) 	& 100 (36) 	& 100 (36) 	\\
640000	& 100 (101) 	& 100 (41)	& 100 (41)  \\
1280000	& 100 (106) 	& 100 (41)  	& 100 (41)  \\
\end{tabular}
\end{table}

 The experiments globally show that  the tuning of    parameter $c$ mainly depends on the value of $q$ even though
 it can be fixed to   small values in all studied cases. Moreover, the presence of non-homogeneous 
 edge-probability function seems to slightly ``help'' the efficiency of the protocol. Intuitively speaking, we believe this is due to the
     presence of fully-random irregularities in the graph topology that  helps the protocol to \emph{break} the  
 symmetry of the initial configuration.

\section{Useful Tools}

\begin{lemma} \label{lem:useful_ineq}
If $x=o(1)$ and $xy = o(1)$ then 
\begin{align*}
\left(1-x\right)^{y}		\geqslant & 1-xy\left(1+ 2 x \right)	 \\
\left(1-x\right)^{y}		\leqslant & 1-xy\left(1-xy\right)				
\end{align*} \end{lemma}

\noindent
We will often use the Chernoff's bounds

\begin{lemma}[Chernoff's Bound.]\label{lemma:cb1}
Let be $X = \sum_{i=1}^n X_i$ where $X_1, \dots, X_n$ are independent 
Bernoulli r.v.s and let be $0<\delta<1$. If  $0 < \mu_1 \leq \mathbf{E}[X]$ 
and $\mu_2 \geq  \Expe{X}$, then it holds that 
\begin{equation} \label{eq:chernoff1}
\mathbf{P} \{ X \leq (1-\delta) \mu_1 \} \leq e^{-\frac{\delta^2}{2}\mu_1}.
\end{equation}

\begin{equation} \label{eq:chernoff2}
\Prob { X \geq (1+\delta) \mu_2 } \leq e^{-\frac{\delta^2}{3}\mu_1}.
\end{equation}
\end{lemma}

\begin{lemma} \label{lem:whpconjunction}
Let $\varphi$ be any poly-logarithm and $E_{0}$, $E_{1}$, ..., $E_{\varphi}$ be events that hold w.h.p.,
then $E_{0}\cap E_{1}\cap...\cap E_{\varphi}$ holds w.h.p. 
\end{lemma}


\section{Proofs of the Protocol's Analysis}

In the sequel, we always analyze the Markovian process in Community $V_1$ since the analysis in the second community is the same.


\subsection{The Source Labeling: Proof of Theorem \ref{thm:endtimephase1}} \label{app:scol}

We  define the following  r.v.s  counting  the labeled  nodes at the end of  Phase 1. 

\begin{itemize}
\item The variable $X_{1}^{v}=1$ iff $v$ gets label $z_{1}$, and the variable $X_{1}=1 + \sum\limits _{ v\neq s_{1}}X_{1}^{v}$ describes the total number of the $z_1$-labeled nodes in $V_1$. 
\item The variable $Y_{1}^{v}=1$ iff $v$ gets label $z_{2}$, and the variable $Y_{1}= \sum\limits _{ v\neq s_{1}}Y_{1}^{v}$ describes the total number of the $z_2$-labeled nodes in $V_1$. 
\end{itemize}

In order to prove the theorem we need the following lemmas.

 \begin{lemma} \label{lem:phase1lowerb}
Let $\tau_1$ be  such that  $p\tau_1 \geqslant  \frac{\log n}n$ and 
$\tau_1=o\left(\frac{1}{p}\right)$. Then, starting from the initial state
 $\left(k_{1}^{(0)} = 1, k_{2}^{(0)}= 1, h_{1}^{(0)}= 0 , h_{2}^{(0)} = 0; E_0 \right)$,   at time step $\tau_{1}$  
 w.h.p. it holds that
\begin{equation*}
\frac{1}{16}np\tau_1 \ \leqslant X_{1},X_2\ \leqslant \ 4np\tau_1 
\end{equation*}
\end{lemma} 
\proof We first bound from below the number of $z_{1}$-labeled
nodes at the end of the   Phase 1. 
Notice that $\Prob{X_{1}^{v}=1}  = \left(1-q\right)^{\tau_{1}} \left(1-\left(1-p\right)^{\tau_{1}}\right)	$ and we can apply Lemma
\ref{lem:useful_ineq} to each factor on the right side, getting:
\[
\Prob{X_{1}^{v}=1}  = \left(1-q\right)^{\tau_{1}} \left(1-\left(1-p\right)^{\tau_{1}}\right)	
 \geqslant  p\tau_{1} \left[\left(1-q\tau_{1}\left(1+{2q}\right)\right)\left(1-p\tau_{1}\right)\right]
 \geqslant  \frac{p\tau_{1}}{2}
\]
where in the last inequality we used 
$\lim_{n \rightarrow \infty} \left[\left(1-q\tau_{1}\left(1+{2q}\right)\right)\left(1-p\tau_{1}\right)\right] =1$.

\noindent The above inequality easily implies that 
\[
\Expe{X_{1}} = 1+ (|V_{1}|-1)\cdot \Prob{X_{1}^{v}=1}
 \ \geqslant \  1+ \left(\frac{n}{2}-1\right)\frac{p\tau_{1}}{2} 
 \ > \  \frac{n}{2} \left(1-\frac{2}{n}\right)\frac{p\tau_{1}}{2},
\] 
that is 
\begin{equation*} 
\Expe{X_{1}} >\frac{np\tau_{1}}{8}
\end{equation*}

\noindent 
Since, by fixing any  initial  state $\left(k_{1}^{(0)},k_{2}^{(0)},h_{1}^{(0)},h_{2}^{(0)}; E_0\right)$,     the r.v.s $X_{1}^{v}$ are
independent, we can apply the Chernoff
Bound (\ref{eq:chernoff1}) with $\delta=\frac{1}{2}$. Then,  
\[
\Prob{X_{1}\leqslant \frac{np\tau_1}{16}} 
\leqslant e^{-\frac{1}{64}np\tau_1}
\] 

\noindent
By hypothesis, we have that  $np\tau_1 \geqslant \log n$, so, w.h.p. it holds that 
\begin{equation*}
X_{1}\geqslant\frac{np\tau_1}{16} 
\end{equation*} 

\noindent  A similar analysis, based on Lemma \ref{lem:useful_ineq} and   Chernoff bound (\ref{eq:chernoff2}),  yields
the stated upper bound on the number of $z_{1}$-labeled
nodes at the end of the   Phase 1, that is, w.h.p. it holds that

\begin{equation*}
X_{1}\leqslant 4np\tau_1 
\end{equation*}
\qed


\begin{lemma} \label{lem:phase1nocom} 
Let    $\tau_{1}\geqslant 1$ be such that $q\tau_1 = O\left(\frac 1{n^{1+\epsilon}}\right)$ for any $\epsilon>0$. Then, 
starting from the initial state  $\left(k_{1}^{(0)} = 1,k_{2}^{(0)}= 1,h_{1}^{(0)}= 0 , h_{2}^{(0)} = 0; E_0 \right)$,  
 at time step $\tau_{1}$  it holds  w.h.p. that $Y_{1}=0$. 
\end{lemma} 
\proof 
A sufficient condition for having $Y_{1}=0$ is that no edge between any node
 in $V_1$ and $s_2$ occurs at any time step of   Phase 1. Hence, by Lemma \ref{lem:useful_ineq}, 
\[ 
\Prob{Y_{1}=0}  \geqslant  \left(1-q\right)^{|V_{1}|\tau_{1}}=\left(1-q\right)^{\frac{n\tau_{1}}{2}}
\geqslant  1-2q|V_{1}|\tau_{1}
\]
Since $q\tau_1=O(\frac{1}{n^{1+ b}})$, the lemma is proved.
\qed

\noindent
  Lemmas \ref{lem:phase1lowerb} and \ref{lem:phase1nocom} easily  imply the 
theorem.  

 
\subsection{ Fast Labeling I: Proof of Theorem \ref{thm:endtimephase2}  } \label{ph2:proof}

We remind that $q = O(p/n^b)$ and consider any positive constant $a$ such that $a<b$.
We   consider   the Markovian   Process when, at the generic step $t$ of this phase, 
 it is  in any    state   satisfying  the following
condition:

\begin{equation} \label{eq:conditionphase2}
k_{1}^{(t)},k_{2}^{(t)}\in\left[\frac{d_1}{16}pn\log n,d_1pn^{1+a }\log n\right] \text{ and }h_{1}^{(t)},h_{2}^{(t)}=0
\end{equation}

 \noindent
For each time step $t$, $\tau_1 < t \leqslant \tau_1+\tau_2$, we define the following binary r.v.s
\begin{itemize}
\item $X_{1}^{v}(t)=1$ iff $v\in V_{1}$ gets label $z_{1}$ at time $t+1$, and $X_{1}(t)=\sum\limits _{v\in V_1}X_{1}^{v}(t)$.
\item $Y_{1}^{v}(t)=1$ iff $v\in V_{1} $ gets label $z_{2}$ at time $t+1$, and $Y_{1}(t)=\sum\limits _{v\in V_1}Y_{1}^{v}(t)$
\end{itemize}

 \noindent
The first lemma provides  tight 
upper and lower bounds on the number of new labeled nodes  after   one   step of the protocol.
The choice of $\tau_2$ will be given later in Theorem \ref{thm:endtimephase2}.

\begin{lemma} \label{lem:phase2stima} 
For $i=1,2$,  it  holds w.h.p.  that
\begin{gather*}
\left(1-\sqrt{\frac{\log n}{k_i^{(t)}}}\right) \left(1-\frac {\polylog n}{ n^{1-a}}\right) \, \frac{np}{2}k_{i}^{(t)}  \leqslant	X_{i}(t) \ \leqslant \left(1+\sqrt{\frac{\log n}{k_i^{(t)}}}\right)\left(1+\frac {\polylog n}{ n^{1-a}} \right)\,  \frac{np}{2}k_{i}^{(t)}	\\
 Y_{i}(t) \ = \ 0.
\end{gather*}
\end{lemma}

\proof 
Observe that $\Prob{Y_{1}(t)=0} $ is lower bounded by the probability that in $E_t$ 
there is no edge between any node in $V_1$ and any node in $V_2$ which is already labeled $z_2$.  
By  the hypothesis  (\ref{eq:conditionphase2}) and the conditions on $p$ and $q$,
we can thus apply Lemma  \ref{lem:useful_ineq}  and get 
\begin{align*}
\Prob{Y_{1}=0} & \geqslant   \left(1-q\right)^{|V_{1}|k_{2}^{(t)}} 
\geqslant   1-2q|V_{1}|k_{2}^{(t)}  \geqslant  1- \frac {\polylog n}{ n^{1-a}}
\end{align*}
proving that w.h.p. $Y_{1}(t)=0$.
Again,  thanks to   Condition  (\ref{eq:conditionphase2}) and the conditions on $p$ and $q$ (Eq. \ref{eq:Cpq}),
we can apply Lemma \ref{lem:useful_ineq} to bound $\Prob{X_{1}^{v}=1}= \left(1-\left(1-p\right)^{k_{1}^{(t)}}\right)\left(1-q\right)^{k_{2}^{(t)}} $. We get
\begin{align*}
\Prob{X_{1}^{v}=1}  & \geqslant k_{1}^{(t)} p \left(1-k_{1}^{(t)}p\right)\left(1-2k_{2}^{(t)}q\right)\geqslant k_{1}^{(t)} p\left(1-\frac {\polylog n}{ n^{1-a}} \right) \\
\Prob{X_{1}^{v}=1}  & \leqslant k_{1}^{(t)} p \left(1+2p\right)\leqslant k_{1}^{(t)} p\left(1+ \frac {\polylog n}{ n^{1-a}}\right) 
\end{align*}
We can thus bound the expected number of new well-labeled nodes $\Expe{X_{1}}   = \left(|V_{1}|-k_{1}^{(t)}\right)\Prob{X_{1}^{v}=1}$:
\begin{align*}\label{eq:phase2lowbX_1}
\frac{np}{2}k_{1}^{(t)}\left(1-\frac {\polylog n}{ n^{1-a}}\right) \leqslant  \Expe{X_{1}} \leqslant \frac{np}{2}k_{1}^{(t)}\left(1+\frac {\polylog n}{ n^{1-a}}\right) 
\end{align*}

\noindent
By applying the Chernoff Bounds (\ref{eq:chernoff1}) and (\eqref{eq:chernoff2})  with   $\delta=\sqrt{\frac{\log n}{k_1}}$, we get

\begin{align*}
\Prob{X_{1}   \leqslant \left(1-\sqrt{\frac{\log n}{k_1^{(t)}}}\right) \left(1-\frac {\polylog n}{ n^{1-a}}\right) \, \frac{np}{2}k_{1}^{(t)}}
 & = e^{-\frac{\log n}{2k_1^{(t)}}\frac{np}{2}k_{1}^{(t)}\left(1-\frac {\polylog n}{ n^{1-a}}\right)} \leqslant   \frac{1}{n^{\frac{1}{3}}}	\\
 \Prob{X_{1}   \geqslant \left(1+\sqrt{\frac{\log n}{k_1^{(t)}}}\right) \left(1+\frac {\polylog n}{ n^{1-a}}\right) \, \frac{np}{2}k_{1}^{(t)}}
 & = e^{-\frac{\log n}{3k_1^{(t)}}\frac{np}{2}k_{1}^{(t)}\left(1-\frac {\polylog n}{ n^{1-a}}\right)} \leqslant   \frac{1}{n^{\frac{1}{3}}}
\end{align*}

\noindent
This implies that, w.h.p., 
\begin{equation*}
\left(1-\sqrt{\frac{\log n}{k_1^{(t)}}}\right) \left(1-\frac {\polylog n}{ n^{1-a}}\right) \, \frac{np}{2}k_{1}^{(t)} \leqslant	X_{1}(t) \ \leqslant \left(1+\sqrt{\frac{\log n}{k_1^{(t)}}}\right)\left(1+\frac {\polylog n}{ n^{1-a}}  \right)\,  \frac{np}{2}k_{1}^{(t)}	 
\end{equation*}
 \qed

\noindent
In what follows,  we will make use of the following function   
\begin{equation*}
F(n,k) = 2\max \left\{ \sqrt{\frac{ \log n}{k}}, \frac {\polylog n}{ n^{1-a}} \right\}
\end{equation*}

\noindent
Let us observe that
 $k_i^{(t+1)} = k_i^{(t)} + X_{i}(t)$. So, 
Lemma \ref{lem:phase2stima}   implies the following recursive bounds 


\begin{lemma} \label{lem:phase2stimaGood}
For $i=1,2$,  it holds w.h.p. that
  
\begin{align*}
\left(1+\frac{np}{2}\right)
\left[1- F(n,k_i^{(t)})\right]k_i^{(t)} & 
\leqslant {k_i^{(t+1)}} \  \leqslant 
\left(1+\frac{np}{2}\right)
\left[1+ F(n,k_i^{(t)})\right]k_i^{(t)}  \\
\mbox{ and } \  & \ {h_i^{(t+1)}} \ =\ 0.   
\end{align*}
\end{lemma}

\smallskip
\noindent
\textbf{Proof of Theorem \ref{thm:endtimephase2}. }
We first analyze the good labeling. 
The idea is to derive the closed formula corresponding to the recurrence relation provided by Lemma \ref{lem:phase2stimaGood} and to analyze it in two different spans of time: in the first span,    we let  $k_1$   increase  enough so that,  in the second span, 
 we can apply a stronger concentration result.
Recall that, thanks to  Theorem \ref{thm:endtimephase1}, Phase 2 starts with the Markovian Process in a state satisfying Condition (\ref{eq:startingCond2}) that implies Condition (\ref{eq:conditionphase2}). 
Moreover, we will fix the final time step $\tau_2$ so that  Condition   (\ref{eq:conditionphase2}) has been holding 
for    \emph{all time steps} of Phase 2: this implies that we can apply  Lemma  \ref{lem:phase2stimaGood} 
for all such steps.  

\noindent
Let $t^*\leqslant  \tau_1$ be defined as
\[ t^\star={\log^{-1}\left(\left(1+\dfrac{np}{2}\right)\left(1	- F(n,k_1^{(\tau_1)})\right)\right)}\log\left(\frac{\log^3 n}{k_1^{(\tau_1)}	}\right)+\tau_1. \]
 Since  $t^*-\tau_1 \in O(\log n)$, thanks to  Lemma \ref{lem:whpconjunction},  
we can  unroll (backward) the recursive relation from time $t^\star$ to time $\tau_1$ and  get 
\begin{equation}\label{eq:F2T_window1}
\left(1+\frac{np}{2}\right)^{t^\star -\tau_1}
\left[1- F(n,k_1^{(\tau_1)})\right]^{t^\star -\tau_1}k_1^{(\tau_1)}  
\leqslant {k_1^{(t^\star)}} \  \leqslant 
\left(1+\frac{np}{2}\right)^{t^\star -\tau_1}
\left[1+ F(n,k_1^{(\tau_1)})\right]^{t^\star -\tau_1}k_1^{(\tau_1)}
\end{equation}

\noindent
We observe  that the value of $k_1^{\tau_1}	\geqslant	\frac{d_1}{16}pn\log n$ can reach  any arbitrarily large constant by tuning the constant $d_1$ in Theorem \ref{thm:endtimephase1}; so,    $F(n,k_1^{(\tau_1)})$ can be made arbitrarily small. From this   fact   
and  Eq. \ref{eq:F2T_window1}, we have that 
$k_1^{(t^\star)}	\in	\left[\log^3 n, \log^{3+\mu}	n	\right]$,
where  $\mu$ can be made arbitrarily small by decreasing $F(n,k_1^{(\tau_1)})$ (i.e. by  increasing $d_1$ in Theorem \ref{thm:endtimephase1}). Notice that, at any time step $t \leqslant t^*$,  Condition (\ref{eq:conditionphase2}) is largely satisfied.

\noindent
We now unroll the recursive relation from time $\tau_2$ to time $t^\star$  and   get 
\begin{equation}\label{eq:F2T_window2}
\left(1+\frac{np}{2}\right)^{\tau_2-t^\star }
\left[1- F(n,k_1^{(t^\star) })\right]^{\tau_2-t^\star }	k_1^{(t^\star) }  
\leqslant {k_1^{(\tau_2)}} \  \leqslant 
\left(1+\frac{np}{2}\right)^{\tau_2-t^\star }
\left[1+ F(n,k_1^{(t^\star) })\right]^{\tau_2-t^\star }	k_1^{(t^\star )}.
\end{equation}

\noindent
We observe  that, with a suitable choice of the positive constant $\phi\in(0,1)$, for
 \[\tau_2=\log^{-1}\left(1+\frac{np}{2}\right)	\log\left(\frac{n^{a}}{\phi \log^3 n}\right)	+	t^\star \]
it holds that  
\begin{align*}
\left[1- F(n,k_1^{(t^\star) })\right]^{\tau_2-t^\star } \geqslant		\phi	\mbox{\hspace{1cm} and \hspace{1cm}}
\left[1+ F(n,k_1^{(t^\star) })\right]^{\tau_2-t^\star }	\leqslant	\frac{1}{\phi} 
\end{align*}

\noindent
By replacing $\tau_2$ into    Eq. \ref{eq:F2T_window2},  with a suitable choice of $\eta \geq \mu$
 (remind that $\mu$ can in turn 
be  made arbitrarily small), we finally get 
$ n^a \leqslant k_1^{(\tau_2)} \leqslant 	  n^a \log^{\eta} n$.
Again, observe  that, for all time steps   $t \leqslant \tau_2$,  Condition (\ref{eq:conditionphase2}) is largely satisfied:
this implies that at each of these steps we were able to apply  Lemma \ref{lem:phase2stimaGood}.

\noindent
As for  the bad labeling, observe  that Lemma \ref{lem:phase2stimaGood} guarantees (w.h.p.)
 ${h}_1^{(t)},{h}_2^{(t)}=0$; then, from Lemma  \ref{lem:whpconjunction},  it holds w.h.p that
${h}_1^{(\tau_2)}=0$ and ${h}_2^{(\tau_2)}=0$.


\subsection{Fast Labeling II: Proof of Theorem \ref{thm:phase3}} \label{ph3:proof}

\noindent
We   consider   the Markovian   Process when, at the generic step $t$ of this phase, 
 it is  in any  state   satisfying  the following
condition
\begin{equation} \label{eq:conditionphase3}
  \mbox{ for }  \ i=1,2  \ : \  \ k_{i}^{(t)} \in\left[n^a,\frac{{n}}{\log^{2} n}\right] \ \text{ and } h_{i}^{(t)} \ = O(  n^{a_2}), \ 
  \mbox{where $a_1< a_2<1$}
  \end{equation}
  
  \noindent
  
  \noindent
For each time step $t$, $\tau_1 < t \leqslant \tau_1+\tau_2$, we again consider  the following binary r.v.s
\begin{itemize}
\item $X_{1}^{v}(t)=1$ iff $v\in V_{1}$ gets label $z_{1}$ at time $t+1$, and $X_{1}(t)=\sum\limits _{v\in V_1}X_{1}^{v}(t)$.
\item $Y_{1}^{v}(t)=1$ iff $v\in V_{1} $ gets label $z_{2}$ at time $t+1$, and $Y_{1}(t)=\sum\limits _{v\in V_1}Y_{1}^{v}(t)$
\end{itemize}

\noindent
In all the next lemmas of this phase, it is  assumed that, 
 at the end of Phase 2, the Markovian Process is in a state satisfying Condition (\ref{eq:conditionphase3})
 (thanks to Theorem \ref{thm:endtimephase2} this holds w.h.p.).

\noindent
We start by providing, with the next lemma, tight upper and lower bounds on the number of the
well-labeled  nodes at a generic step of Phase 3.


\begin{lemma}\label{lm:step_k3}
A constant  $\zeta>0$ exists such that, for $i=1,2$,  it holds  w.h.p. that
\begin{align}
\left(1+\frac{np}2 \right)\left(1-\frac{\zeta }{\log n}\right)k_i^{(t)} & 
\leqslant \  k_i^{(t+1)} 
\leqslant \ \left(1+\frac{np}2 \right)\left(1+\frac{\zeta}{\log n}\right)k_i^{(t)}         \label{eq:phase2lem1result1} 
\end{align}
\end{lemma}

\skproof 
By neglecting the contribution of $h_2$,  from the facts 
 $pk_1^{(t)},qk_2^{(t)},ph_1^{(t)} = o(1) $ and   Lemma \ref{lem:useful_ineq},  we have that 
\begin{eqnarray*} 
\Prob{X_1^v=1} & \geqslant & \left(1-\left(1-p\right)^{k_1^{(t)}}\right)\left(1-q\right)^{k_2^{(t)}} \left(1-p\right)^{h_1^{(t)}} \\
& \geqslant &	pk_1^{(t)}\left(1-p{k_1^{(t)}}\right)	\left(1-2q{k_2^{(t)}}\right) \left(1-2p{h_1^{(t)}}\right).
\end{eqnarray*}

\noindent
Observe that   

\[ \left(1-p{k_1^{(t)}}\right)	\left(1-2q{k_2^{(t)}}\right) \left(1-2p{h_1^{(t)}}\right) \ \geqslant \ \left(1-\Theta\left(\frac 1 {\log^2n}\right)\right) \]
then
\begin{align}
\Prob{X_1^v=1}	\geqslant	pk_1^{(t)} \left(1-\Theta\left(\frac 1 {\log^2n}\right)\right) \label{eq:F3L2lowerX}
\end{align} 

\noindent We now provide an upper  bound on  $\Prob{X_1^v=1}$. From  the Union Bound and Lemma \ref{lem:useful_ineq}, we get
\begin{align}
\Prob{X_1^v=1}	\leqslant	pk_1^{(t)}\left(1+2{p}\right)	+	qh_2^{(t)}\left(1+2{q}\right)	\leqslant	pk_1^{(t)}\left(1+\Theta \left(\frac 1{n^{1-a_2}}\right)\right).\label{eq:F3L2upperX} 
\end{align}

\noindent
As usual, we exploit Eq.s \ref{eq:F3L2lowerX} and \ref{eq:F3L2upperX} to bound 
the expectation 
\[ \Expe{X_1}=\left(|V_1|-(k_1^{(t)}+h_1^{(t)})\right)\cdot \Prob{X_1^v=1} \] 
For some constant $\tilde \zeta>0$, w.h.p. it thus holds that
\begin{equation*} 
\frac{np}{2}k_1^{(t)}	\left(1-\frac{\tilde \zeta}{\log n}\right)	
\leqslant	\Expe{X_1}	\leqslant
\frac{np}{2}k_1^{(t)}	\left(1+\frac{\tilde \zeta}{\log n}\right) 
\end{equation*}

\noindent
We can   use the Chernoff Bounds (\ref{eq:chernoff1} and \ref{eq:chernoff2} with $\delta={1}/{\log n}$),
  to get that, for some constant ${\zeta}>0$, w.h.p. 
\begin{equation}\label{eq:F3L1upperX1}
\frac{np}{2}k_1^{(t)}	\left(1-\frac{\zeta}{\log n}\right)		
\leqslant	X_1	\leqslant	
\frac{np}{2}k_1^{(t)}\left(1+\frac{ \zeta}{\log n}\right)
\end{equation}
 From the above inequality,
it follows that 
\begin{align*} 
\left(1+\frac{np}{2}\right)\left(1-\frac{np}{2+np}\frac{\zeta}{\log n}\right)k_1^{(t)}	
\leqslant	k_1^{(t+1)}	
\leqslant	\left(1+\frac{np}{2}\right)\left(1+\frac{np}{2+np}\frac{\zeta}{\log n}\right)k_1^{(t)},
\end{align*}
Since $\frac{np}{2+np}{\zeta}$ is bounded by a constant, for the sake of simplicity we can just re-define $\zeta$ as any fixed 
 constant such that 
\begin{align*} 
\left(1+\frac{np}{2}\right)\left(1-\frac{\zeta}{\log n}\right)k_1^t	
\leqslant	k_1^{t+1}	
\leqslant	\left(1+\frac{np}{2}\right)\left(1+\frac{\zeta}{\log n}\right)k_1^t 
\end{align*}
\qed

\noindent
  Lemma  \ref{lm:step_k3} implies the following properties of the well-labeling for any state within Phase 3 (including 
   the final one at time $\tau_3$).

\begin{lemma} \label{lm:k3}
A constant  $\zeta>0$ exists such that, for $i=1,2$, it holds w.h.p. that 
\[ \left(1+\frac{np}{2}\right)^{t+1-\tau_2}	\left(1-\frac{\zeta}{\log n}\right)^{t+1-\tau_2}	k_1^{(\tau_2)}	
\leqslant	k_1^{(t+1)}	
\leqslant	\left(1+\frac{np}{2}\right)^{t+1-\tau_2}	\left(1+\frac{\zeta}{\log n}\right)^{t+1-\tau_2}	k_1^{(\tau_2)} \]
\end{lemma}

\skproof 
Since Lemma \ref{lm:step_k3} holds as long as $k_i	\leqslant	\frac{n}{ \log^{2} n }$, from Lemma \ref{lem:whpconjunction} we get
that 
the statement holds w.h.p. by applying the same unrollement  argument shown in the previous phase.
\qed

 \noindent
 We now exploit the above lemma to   provide a bound on the number of bad-labeled nodes at the end of Phase 3.
 
\begin{lemma} \label{lm:h3}
For any positive constant  $\gamma$, a constant $a_1$, with  $ 1 -a<a_1<a_2$, can be fixed so that
by choosing  the final time step of Phase 3
\[ \tau_3 =  \frac 1{\log\left(1+\left(\frac{np}2\right)\right)} \log\left( \frac{n^{1-a}} {\gamma \log^{3}{n}}\right) +\tau_2 , \]
   it holds w.h.p.  that, for $i=1,2$ and for all $t \leqslant \tau_3$, 
${h}_i^{(t)}\  \leqslant \ {n^{a_1}}$.
\end{lemma}
 
\skproof 
 In order to bound   the rate  of $h_1^{(t)}$, we consider the r.v.  $Y_1^v$ when the Markovian Process is 
  in a generic state satisfying Condition (\ref{eq:conditionphase3}). Thanks to Theorem \ref{thm:endtimephase2}, 
  we know this (largely) holds for the first step of Phase 3 and, by the choice of $\tau_3$, we will see this  (w.h.p.) holds 
  for all $t \leqslant \tau_3$ by 
    induction.

\noindent By neglecting the contribution of $k_2$, we have that 
\[ \Prob{Y_1^v=1}	\geqslant	\left(1-\left(1-p\right)^{h_1^{(t)}}\right)	\left(1-q\right)^{h_2^{(t)}} \left(1-p\right)^{k_1^{\left(t\right)}} \]

\noindent
 Since $ph_1^{(t)},qh_2^{(t)},pk_1^{(t)}  = o( 1)$, we can apply Lemma \ref{lem:useful_ineq}   to each factor on the right side, thus  obtaining
\begin{align*}\label{eq:f3lowerPYpartial}
 \Prob{Y_1^v=1}	&\geqslant	ph_1^t\left(1-p{h_1^{(t)}}\right)	\left(1-2q{h_2^{(t)}}\right) \left(1-2p{k_1^{(t)}}\right) \\ 
 & \geqslant		ph_1^t\left(1-\Theta\left(\frac 1{\log^2n}\right)\right)
\end{align*} 

\noindent
We now provide an upper bound to  $\Prob{Y_1^v=1}$.
By the Union Bound and Lemma \ref{lem:useful_ineq} we get
\begin{align*}
\Prob{Y_1^v=1}	&\leqslant	\left(1-\left(1-p\right)^{h_1^{(t)}}\right)	+\left(1-\left(1-q\right)^{k_2^{(t)}}\right)\\
				&\leqslant	ph_1^{(t)}\left(1+{2p}\right)				+qk_2^{(t)}\left(1+2q\right)\\
				&\leqslant	\left(ph_1^{(t)}				+qk_2^{(t)}\right)\left(1+{2p}\right) 
\end{align*}

\noindent
As for the expected value of new bad-labeled nodes,   for some constant $\zeta>0$, it holds that 
\begin{equation}
\frac{np}{2}h_1^{(t)}	\left(1-\frac{\zeta}{\log n}\right)
\leqslant	\Expe{Y_1}	\leqslant	
\left(\frac{np}{2}h_1^{(t)}	+   \frac{nq}{2}k_2^{(t)}\right)    \left(1+\frac 1{\log n}\right) \label{eq:F3Lh_expe_bound}
\end{equation}

\noindent From the  Chernoff Bound and Eq.   (\ref{eq:F3Lh_expe_bound}), it follows that w.h.p. $h_1$ 
will not ``jump'' from a sublogarithmic value to a polynomial one: in other words, in the first time that $T$ will be at least $\log^3 n$,
 we have that $h_1^{T} = O(\polylog n)$.

\noindent Hence, again from the  Chernoff Bound and Eq.   (\ref{eq:F3Lh_expe_bound}),
setting  $\delta=\sqrt {\frac{\log n}{h_1^{(t)}}}$, we see that for each $t \geqslant T$ in Phase 3 w.h.p. we have 
\begin{equation}
Y_1 \leqslant \left(\frac{np}{2}h_1^{(t)}	+\frac{k_2^{(t)}}{2n^\alpha}\right) \left(1+\frac 2{\log n}\right) \label{Y_bound}
\end{equation}

In Eq. (\ref{Y_bound}), we can bound the term $\frac{k_2^{(t)}}{2n^\alpha}$ using Lemma \ref{lm:k3} and Theorem \ref{thm:endtimephase2}, obtaining for some positive constant $c$
\begin{equation*}
 \frac{k_2^{(t)}}{2n^\alpha} \leqslant \left(1+\frac{np}{2}\right)^{t-\tau_2} \frac{\left(1+\frac{\zeta}{\log n}\right)^{t-\tau_2}	k_1^{(\tau_2)} }{2n^\alpha} \leqslant \left(1+\frac{np}{2}\right)^{t-\tau_2} \cdot c
\end{equation*}

Therefore we can use Eq. (\ref{Y_bound}) to get that w.h.p.
\begin{equation}
 h_1^{(t+1)}=h_1^t+Y_1 \leqslant \left( \left(1+\frac{np}{2}\right)h_1^{(t)}	+\left(1+\frac{np}{2}\right)^{t-\tau_2} \cdot c \right) \left(1+\frac 2{\log n}\right) \label{eq:h_recursion}
\end{equation}

Hence unrolling $h_1^{(t)}$ until time $T$, for some positive constants $c_1$ and $c_2$, Eq. ( \ref{eq:h_recursion}) becomes (keeping high probability thanks to Lemma \ref{lem:whpconjunction}) 
\begin{align*}
 h_1^{t+1} &\leqslant \left(1+\frac{np}{2}\right)^{t+1-T} \cdot \left(1+\frac 2{\log n}\right)^{t+1-T} \cdot h_1^{(T)} + c \cdot \left(1+\frac{np}{2}\right)^{t-\tau_2} \cdot \sum_{i=T}^{t}  \left(1+\frac 2{\log n}\right)^{t+1-i} \label{eq:Y_well_written_with_T}\\
 &\leqslant c_1 \left(1+\frac{np}{2}\right)^{t+1-T} h_1^{(T)} + c_2 \log n \left(1+\frac{np}{2}\right)^{t-\tau_2} 
\end{align*}
and the last side turns out to be $O(n^{1-a} \cdot \polylog n)$ when $t+1=\tau_3$, proving the lemma. \qed

\medskip

\noindent
{\bf Proof of Theorem \ref{thm:phase3}.}

\noindent
The  bound claimed for   $h_i$   follows  from  Lemma \ref{lm:h3}, and the bounds claimed for $k_i$ follow from Lemma \ref{lm:k3} for $t=\tau_3$, thus  proving  Theorem \ref{thm:phase3}   \qed

\subsection{Dealing with   Stochastic Dependence} \label{app:stocdep}

Here we prove the  properties yielded by Procedure \orlink\  claimed   in Phase 4 of Section \ref{ssec:restcase}. Since we are considering a generic time step, we  omit its index, thus $E=E_t$. We denote with $E^D$ the set of directed edges constructed by  Procedure \orlink; then, when writing $(u,v)\in E^D$ we are assuming that $(u,v)$ is a directed edge.

\begin{lemma}
$(\mathcal{V}, E^D)$ is distributed as a directed $G_{|\mathcal{V}|,(1-\sqrt{1-p})}$, up to a total variation error of $O\left(\frac 1{n^{c-2}}\right)$ for an arbitrary positive $c$ set in the subprocedure.
\end{lemma}

\proof
In what follows, the approximations denoted by ``$\approx$'' consist in dropping off factors of order $O\left(\frac 1{n^c}\right)$, where $c$ is given by the range from whom the numbers $M^{\cdot}$ are sampled, that is from 1 to $n^c$ (for the sake of simplicity in our protocol we set $c=3$). 

\noindent In order to study the distribution of $(\mathcal{V}, E^D)$, observe that for a given node $u$ respect to a node $v$ it holds
\begin{align*}
&\Prob{C \in \{0,1\} \,\middle|\, (u,v)\in E} = \\ 
&= \Prob{M^{(u,v)} > M^{(v,u)}}\cdot \Prob{C \in \{0,1\} \,\middle|\, M^{(u,v)} > M^{(v,u)}}+\\
&+ \Prob{M^{(u,v)} < M^{(v,u)}}\cdot \Prob{C \in \{0,1\} \,\middle|\, M^{(u,v)} < M^{(v,u)}} + O\left(\Prob{M^{(u,v)} = M^{(v,u)}}\right)\approx\\
&\approx \frac 12 \cdot \Prob{C \in \{0,1\} \,\middle|\, M^{(u,v)} > M^{(v,u)}}+ \frac 12 \cdot \Prob{C \in \{0,1\} \,\middle|\, M^{(u,v)} < M^{(v,u)}}  =\frac{1-\sqrt{1-p}}{p} 
\end{align*}
From the preceding calculation for any pair of nodes $u,v \in \mathcal{V}$, it follows
\begin{align*}
\Prob{(u,v)\in E^D} &= \Prob{(u,v)\in E}\cdot \Prob{C \in \{0,1\} \,\middle|\, (u,v)\in E} \approx \\
&\approx \Prob{(u,v)\in E} \cdot \frac{1-\sqrt{1-p}}{p}  = 1-\sqrt{1-p}
\end{align*}
Now we have to shows that the edges of $E^D$ are independent. Since r.v.s that are functions of independent r.v.s are themselves independent, notice that two given edges $(u,v),(w,z)\in E^D$, that have one or zero nodes in common, are independent because they are built on independent edges of the $\sdG(n,\tilde p,\tilde q)$ graph.
It remains to verify that the edges $(u,v),(v,u)\in E^D$ (that are built on the same edge), are independent. By direct calculation
\begin{align*}
\Prob{(u,v)\in E^D  \,\middle|\,  (v,u)\in E^D} &= \frac{ \Prob{(u,v),  (v,u)\in E^D}} {\Prob{  (v,u)\in E^D}}=  \frac{\Prob{(u,v)\in E \text{ and } C = 0}}{\Prob{  (v,u)\in E^D}} \approx \\
&\approx \frac{(1-\sqrt{1-p})^2}{1-\sqrt{1-p}}=1-\sqrt{1-p} \approx \Prob{  (u,v)\in E^D}
\end{align*} concluding the proof.

\qed


\subsection{Controlled Saturation: Proof of Theorem \ref{thm:kh4}} \label{ph4:proof}

We  define the following  r.v.s  counting  the labeled  nodes at the end of each window of Phase 4. 

\begin{itemize}
\item The variable $X_{1}^{v}=1$ iff $v$ gets label $z_{1}$, and the variable $X_{1}=1 + \sum\limits _{ v\neq s_{1}}X_{1}^{v}$ describes the total number of the $z_1$-labeled nodes in $V_1$. 
\item The variable $Y_{1}^{v}=1$ iff $v$ gets label $z_{2}$, and the variable $Y_{1}= \sum\limits _{ v\neq s_{1}}Y_{1}^{v}$ describes the total number of the $z_2$-labeled nodes in $V_1$. 
\end{itemize}

\noindent Observe that because of Procedure \orlink{} the edge probabilities change, however to simplify notation we keep using $p$ and $q$ for the new edge probabilities.

 \begin{lemma} \label{lm:h4}
For any  constant $c_4$, at time step $\tau_4 = 3 T_4+ \tau_3 = 3 c_4 \log n + \tau_3$, the Markovian  Process is  w.h.p. in a state such that 
 $h_i^{\tau_4} = O(n^{a_1} \ \polylog n)$, for $i=1,2$.
 \end{lemma}
 
 \skproof
 Let $h_1=h_1^{(\tau_3)}$ and $k_2 = k_2^{(\tau_3)}$. At the end of the first window, for any node $v \in V_1$, it holds that
 \begin{align*} \Prob{Y_1^v =1} \  \leqslant & \ \left(1-\left( (1-p)^{h_1} (1-q)^{k_2}\right)^{T_4}\right) \\
 \leqslant & \ 1 - e^{-T_4(ph_1 + qk_2)} \\
 \leqslant &   \ T_4(ph_1 + qk_2) 
 \end{align*}
 
 \noindent
So, since it holds $a_1>1-a>1-b$, we get 
$\Expe{Y_1} \  \leqslant \ 2 \ nph_1T_4 \ =  \ O\left( n^{a_1} \ \polylog n\right)$. \\
 By repeating the same reasoning for the other 2 windows and by applying the  Chernoff bound, the thesis follows.
 
 \qed

\smallskip
\noindent
\textbf{Proof of Theorem \ref{thm:kh4}.}
For the sake of brevity, we define $k_1 = k_1^{(\tau_3)} $, $k_2 = k_2^{(\tau_3)} $, and $h = h^{(\tau_3)}_1$.\\
Let us consider a node $v \in V_1$ at the end of the first time window of Phase 4.
For    some constant $\gamma > 0$, it holds that
\begin{align*}
\Prob{X_1^v = 1} \ \geqslant & \ \left(  1 - (1-p)^{k_1T_4}\right) (1-p)^{h_1T_4} (1-q)^{k_2T_4} \\
   & \geqslant \left(1- \frac{\gamma}{\log^{1-\eta}n} \right) pk_1T_4.
   \end{align*}
   
   \noindent
Since $\frac 1n \leqslant p \leqslant \frac{\log n}n$, by computing the expected value of the sum of all $X_1^v$'s and by applying the Chernoff bound, we get
that the number of well-labeled nodes in $V_1$ at the end of the first time window of Phase 4 is w.h.p.
   
   \[ k_1^{T_4} \ \geqslant \ d_4 \frac{n}{\log^2 n},    \]
   where $d_4 = d_4(c_4)$ is a positive constant that can be made arbitrarily large by increasing $c_4$ in $T_4= c_4 \log n$.

   \noindent
   We have thus shown that, after the first window, the number of well labeled nodes inside each community is increased
   by a  factor $d_4 \log n$. 
   We can then repeat the same analysis for the second and the third windows (which are  necessary when $p=o(\log n/n)$).
   Let us consider the sparsest case  $p=1/n$ (the other cases are easier).
   In this case, at the end of the third window, it can be easily verified that:
   
  \begin{align*}
   \Prob{X_1^v = 1} \ \geqslant & \ \left(  1 - (1-p)^{\frac{n}{\log n}T_4}\right) (1-p)^{( n^{a_1}  \polylog n) \  T_4 } (1-q)^{n T_4} \\
   & \geqslant \left(1- \frac{1}{n^{\epsilon}} \right) \left( 1- e^{-c_4} \right)    
   \end{align*} 
   The last bound can be thus made arbitrarily close to 1 by increasing the constant $c_4$. Hence,
 w.h.p.  
   
     \[ k_1^{(\tau_4)} \ \geqslant \alpha n   \]
     where      constant $\alpha$   can be made arbitrarily close to $1$  by suitably choosing  
     the constant $c_4$ in $T_4 = c_4 \log n$.

\noindent
As for the bad labeling, the thesis follows from  Lemma \ref{lm:h4}.

\qed

\subsection{Majority Rule: Proofs of Theorem \ref{thm:F5all}} \label{ph5:proof}
  \skproof
 Let us consider a node $u \in V_1$ and, for every time step $t$ of   Phase 5, define the    r.v. $X^u_t$  
 counting the number of its $z_1$-labeled neighbors and the r.v. $Y^u_t$ 
 counting the number of its $z_2$-labeled neighbors in $E_t$.
 Then, define  the two sums 
 
\[ X_u \  =  \ \sum_{t \in [\tau_4 +1,\ldots, \tau_5] } X_t^u \  \ \mbox{ and } \  \ Y_u = \sum_{t \in [\tau_4+1,\ldots, \tau_5] } Y_t^u \]

\noindent
Let us also define the subset 
\[ G^{\tau_4} =
 \{ v \in V_1 \ | \ v  \ \mbox{ is $z_1$-labeled at time } \  \tau_4 \} \]
 Thanks to Condition  \ref{eq::f4final} (with constant $\alpha = 3/4$), it holds  that 

\begin{equation*} 
| G^{\tau_4} | \ \geqslant \ \frac 34 \,  |V_1| \  =  \ \frac 38 \,  n
\end{equation*}
 
 \noindent
From the above inequality,
the expected values of r.v.s $X_u$ and $Y_u$  can be easily bound  as follows
 
 \[ \Expe{X_u}  \
  \geqslant \ \sum_{t \in [\tau_4+1,\ldots, \tau_5] } \sum_{v \in G^{\tau_4}} \Prob{(u,v) \in E_t} \geqslant \frac 38 \, pn \tau_5 \ , \ \mbox{ and }  \]

\[  \Expe{Y_u}  \ \leqslant \ \sum_{t \in [\tau_4+1,\ldots, \tau_5] } \left( \sum_{v \not\in G^{\tau_4}} \Prob{(u,v) \in E_t} + \sum_{v \in V_2}  \Prob{(u,v) \in E_t} \right)  \leqslant \ \frac 17 \, pn \tau_5 \]
 
\noindent
Finally, observe that $X_u$ and $Y_u$ are sums of independent binary r.v.s (thanks to Procedure \orlink). Since $p \geqslant 1/n$ and $\tau_5 = \tau_4 + c_5 \log n$, we can thus  choose a suitable constant
$c_5>0$ and apply the Chernoff bound
to get the thesis. 

 \qed
 

\newpage


 
\begin{thebibliography}{10}

\bibitem{AKL08}
Chen Avin, Michal Kouck{\`y}, and Zvi Lotker.
\newblock How to explore a fast-changing world (cover time of a simple random
  walk on evolving graphs).
\newblock In {\em Automata, Languages and Programming}, pages 121--132.
  Springer, 2008.

\bibitem{BC09}
Michael~J. Barber and John~W. Clark.
\newblock Detecting network communities by propagating labels under
  constraints.
\newblock {\em Phys. Rev. E}, 80:026129, Aug 2009.

\bibitem{BCF09}
Herv{\'e} Baumann, Pierluigi Crescenzi, and Pierre Fraigniaud.
\newblock Parsimonious flooding in dynamic graphs.
\newblock In {\em Proceedings of the 28th ACM symposium on Principles of
  distributed computing}, PODC '09, pages 260--269, New York, NY, USA, 2009.
  ACM.

\bibitem{BLMCH06}
S.~Boccaletti, V.~Latora, Y.~Moreno, M.~Chavez, and D.-U. Hwang.
\newblock Complex networks: Structure and dynamics.
\newblock {\em Physics Reports}, 424(4-5):175 -- 308, 2006.

\bibitem{B87}
Ravi~B. Boppana.
\newblock Eigenvalues and graph bisection: An average-case analysis.
\newblock In {\em Proceedings of the 28th Annual Symposium on Foundations of
  Computer Science}, SFCS '87, pages 280--285, Washington, DC, USA, 1987. IEEE
  Computer Society.

\bibitem{BDG+07}
Ulrik Brandes, Daniel Delling, Marco Gaertler, Robert G\"orke, Martin Hoefer,
  Zoran Nikoloski, and Dorothea Wagner.
\newblock On modularity clustering.
\newblock {\em IEEE Transactions on Knowledge and Data Engineering},
  20(2):172--188, 2008.

\bibitem{BCLS87}
Thang~Nguyen Bui, F.~Thomson Leighton, Soma Chaudhuri, and Michael Sipser.
\newblock Graph bisection algorithms with good average case behavior.
\newblock {\em Combinatorica}, 7(2):171--191, June 1987.


\bibitem{CG10}  Gennaro Cordasco and
               Luisa Gargano.
               \newblock 
 Label propagation algorithm: a semi-synchronous approach.
 \newblock 
   \emph{IJSNM}, 1(1), 3-26, 2012.  {http://dx.doi.org/10.1504/IJSNM.2012.045103}.
  
\bibitem{CHCDSG07}
A.~Chaintreau, Pan Hui, J.~Crowcroft, C.~Diot, R.~Gass, and J.~Scott.
\newblock Impact of human mobility on opportunistic forwarding algorithms.
\newblock {\em Mobile Computing, IEEE Transactions on}, 6(6):606--620, 2007.

\bibitem{CMMD07}
Augustin Chaintreau, Abderrahmen Mtibaa, Laurent Massoulie, and Christophe
  Diot.
\newblock The diameter of opportunistic mobile networks.
\newblock In {\em Proceedings of the 2007 ACM CoNEXT conference}, CoNEXT '07,
  pages 12:1--12:12, New York, NY, USA, 2007. ACM.

\bibitem{CMMPS08}
Andrea~E.F. Clementi, Claudio Macci, Angelo Monti, Francesco Pasquale, and
  Riccardo Silvestri.
\newblock Flooding time in edge-markovian dynamic graphs.
\newblock In {\em Proceedings of the twenty-seventh ACM symposium on Principles
  of distributed computing}, PODC '08, pages 213--222, New York, NY, USA, 2008.
  ACM.

\bibitem{CMPS09}
Andrea~E.F. Clementi, Angelo Monti, Francesco Pasquale, and Riccardo Silvestri.
\newblock Information spreading in stationary markovian evolving graphs.
\newblock In {\em Parallel \& Distributed Processing, 2009. IPDPS 2009. IEEE
  International Symposium on}, pages 1--12. IEEE, 2009.

\bibitem{CK01}
Anne Condon and Richard~M Karp.
\newblock Algorithms for graph partitioning on the planted partition model.
\newblock {\em Random Structures and Algorithms}, 18(2):116--140, 2001.

\bibitem{DDDA05}
Leon Danon, Albert Diaz-Guilera, Jordi Duch, and Alex Arenas.
\newblock Comparing community structure identification.
\newblock {\em Journal of Statistical Mechanics: Theory and Experiment},
  2005(09):P09008, 2005.

\bibitem{DF89}
M.E Dyer and A.M Frieze.
\newblock The solution of some random np-hard problems in polynomial expected
  time.
\newblock {\em Journal of Algorithms}, 10(4):451 -- 489, 1989.

\bibitem{EP05}
Nathan Eagle and Alex Pentland.
\newblock Reality mining: sensing complex social systems.
\newblock {\em Personal and ubiquitous computing}, 10(4):255--268, 2006.

\bibitem{EK10}
David Easley and Jon Kleinberg.
\newblock {\em Networks, crowds, and markets}, volume~8.
\newblock Cambridge Univ Press, 2010.

\bibitem{FL02}
G.W. Flake, S.~Lawrence, C.L. Giles, and F.M. Coetzee.
\newblock Self-organization and identification of web communities.
\newblock {\em Computer}, 35(3):66--70, 2002.

\bibitem{GN02}
M.~Girvan and M.~E.~J. Newman.
\newblock Community structure in social and biological networks.
\newblock {\em Proceedings of the National Academy of Sciences},
  99(12):7821--7826, 2002.

\bibitem{HRT07}
Mark~S Handcock, Adrian~E Raftery, and Jeremy~M Tantrum.
\newblock Model-based clustering for social networks.
\newblock {\em Journal of the Royal Statistical Society: Series A (Statistics
  in Society)}, 170(2):301--354, 2007.

\bibitem{HLL83}
Paul~W. Holland, Kathryn~Blackmond Laskey, and Samuel Leinhardt.
\newblock Stochastic blockmodels: First steps.
\newblock {\em Social Networks}, 5(2):109 -- 137, 1983.

\bibitem{HYCC07}
Pan Hui, Eiko Yoneki, Shu~Yan Chan, and Jon Crowcroft.
\newblock Distributed community detection in delay tolerant networks.
\newblock In {\em Proceedings of 2nd ACM/IEEE international workshop on
  Mobility in the evolving internet architecture}, page~7. ACM, 2007.

\bibitem{JS98}
Mark Jerrum and Gregory~B Sorkin.
\newblock The metropolis algorithm for graph bisection.
\newblock {\em Discrete Applied Mathematics}, 82(1):155--175, 1998.

\bibitem{KLV07}
Thomas Karagiannis, Jean-Yves Le~Boudec, and Milan Vojnovi{\'c}.
\newblock Power law and exponential decay of inter contact times between mobile
  devices.
\newblock In {\em Proceedings of the 13th annual ACM international conference
  on Mobile computing and networking}, pages 183--194. ACM, 2007.

\bibitem{KPS13}
Kishore Kothapalli, Sriram~V Pemmaraju, and Vivek Sardeshmukh.
\newblock On the analysis of a label propagation algorithm for community
  detection.
\newblock In {\em Distributed Computing and Networking}, pages 255--269.
  Springer, 2013.
  
  \bibitem{LHLC09} I.X.Y.  Leung, P. Hui, P. Li\'o, and J. Crowfort.
  \newblock Towards real-time community detection algorithms in large networks.
  \newblock \emph{Phys. Rev. E.} 79(6), 2009.


\bibitem{LM10}
Xin Liu and Tsuyoshi Murata.
\newblock Advanced modularity-specialized label propagation algorithm for
  detecting communities in networks.
\newblock {\em Physica A: Statistical Mechanics and its Applications},
  389(7):1493--1500, 2010.

\bibitem{LN04}
David Lusseau and M.~E.~J. Newman.
\newblock Identifying the role that animals play in their social networks.
\newblock {\em Proceedings of the Royal Society of London. Series B: Biological
  Sciences}, 271(Suppl 6):S477--S481, 2004.

\bibitem{MRRS12}
Francesca Martelli, M~Elena~Renda, Giovanni Resta, and Paolo Santi.
\newblock A measurement-based study of beaconing performance in ieee 802.11 p
  vehicular networks.
\newblock In {\em INFOCOM, 2012 Proceedings IEEE}, pages 1503--1511. IEEE,
  2012.

\bibitem{M01}
Frank McSherry.
\newblock Spectral partitioning of random graphs.
\newblock In {\em Foundations of Computer Science, 2001. Proceedings. 42nd IEEE
  Symposium on}, pages 529--537. IEEE, 2001.

\bibitem{MNS12}
E.~{Mossel}, J.~{Neeman}, and A.~{Sly}.
\newblock {Stochastic Block Models and Reconstruction}.
\newblock {\em ArXiv e-prints}, February 2012.

\bibitem{N06}
M.~E.~J. Newman and M.~Girvan.
\newblock Finding and evaluating community structure in networks.
\newblock {\em Phys. Rev. E}, 69:026113, Feb 2004.

\bibitem{N02}
Mark~EJ Newman.
\newblock Spread of epidemic disease on networks.
\newblock {\em Physical review E}, 66(1):016128, 2002.

\bibitem{NG04}
Mark~EJ Newman and Michelle Girvan.
\newblock Finding and evaluating community structure in networks.
\newblock {\em Physical review E}, 69(2):026113, 2004.

\bibitem{PSSL09}
Chuanjun Pang, Fengjing Shao, Rencheng Sun, and Shujing Li.
\newblock Detecting community structure in networks by propagating labels of
  nodes.
\newblock In {\em Advances in Neural Networks--ISNN 2009}, pages 839--846.
  Springer, 2009.

\bibitem{RAK07}
Usha~Nandini Raghavan, R\'eka Albert, and Soundar Kumara.
\newblock Near linear time algorithm to detect community structures in
  large-scale networks.
\newblock {\em Phys. Rev. E}, 76:036106, Sep 2007.

\bibitem{RSMOB02}
Erzs{\'e}bet Ravasz, Anna~Lisa Somera, Dale~A Mongru, Zolt{\'a}n~N Oltvai, and
  A-L Barab{\'a}si.
\newblock Hierarchical organization of modularity in metabolic networks.
\newblock {\em science}, 297(5586):1551--1555, 2002.

\bibitem{SN97}
Tom~AB Snijders and Krzysztof Nowicki.
\newblock Estimation and prediction for stochastic blockmodels for graphs with
  latent block structure.
\newblock {\em Journal of Classification}, 14(1):75--100, 1997.

\bibitem{SJP08}
Thrasyvoulos Spyropoulos, Apoorva Jindal, and Konstantinos Psounis.
\newblock An analytical study of fundamental mobility properties for
  encounter-based protocols.
\newblock {\em International Journal of Autonomous and Adaptive Communications
  Systems}, 1(1):4--40, 2008.

\bibitem{TL09}
P-U Tournoux, J{\'e}r{\'e}mie Leguay, Farid Benbadis, Vania Conan, M~Dias
  De~Amorim, and John Whitbeck.
\newblock The accordion phenomenon: Analysis, characterization, and impact on
  dtn routing.
\newblock In {\em INFOCOM 2009, IEEE}, pages 1116--1124. IEEE, 2009.

\bibitem{V11}
Milan Vojnovic and Alexandre Proutiere.
\newblock Hop limited flooding over dynamic networks.
\newblock In {\em INFOCOM, 2011 Proceedings IEEE}, pages 685--693. IEEE, 2011.

\bibitem{WCdeA11}
John Whitbeck, Vania Conan, and Marcelo~Dias de~Amorim.
\newblock Performance of opportunistic epidemic routing on edge-markovian
  dynamic graphs.
\newblock {\em Communications, IEEE Transactions on}, 59(5):1259--1263, 2011.

\bibitem{YHC08}
Eiko Yoneki, Pan Hui, and Jon Crowcroft.
\newblock Wireless epidemic spread in dynamic human networks.
\newblock In {\em Bio-Inspired Computing and Communication}, pages 116--132.
  Springer, 2008.

\end{thebibliography}
 \end{document}